\documentclass{article}

\usepackage{graphicx}
\usepackage[a4paper, total={7in, 10in}]{geometry}
\usepackage{caption}
\usepackage{subcaption}
\usepackage{comment}
\usepackage{amsmath}
\usepackage{amssymb}
\usepackage{array}
\usepackage{xcolor}
\usepackage{float}
\usepackage{paralist}
\usepackage{authblk}
\usepackage{multirow}
\usepackage{booktabs}
\usepackage[backend=biber,doi=false,isbn=false,url=false,eprint=false]{biblatex} 
\usepackage[title,toc,page,header]{appendix}

\title{Compressible single-phase flows in heated and cooled ducts}

\author[1]{S. Schropff\thanks{solene.schropff@univ-amu.fr}}
\author[1]{F. Petitpas\thanks{fabien.petitpas@univ-amu.fr}}
\author[1]{E. Daniel\thanks{eric.daniel@univ-amu.fr}}
\affil[1]{Aix Marseille Univ, CNRS, IUSTI, Marseille, France}

\begin{document}

\include{figures}

    \maketitle
    \begin{abstract}
Analytical/quasi-analytical solutions are proposed for a steady, compressible, single-phase flow in a rectilinear duct subjected to heating followed by cooling. The flow is driven by the pressure ratio between an upstream tank and a variable outlet pressure. The article proposes a methodology to determine the full flow behavior, as a function of pressure ratio and heat-flux distribution. Following an analogy done with the study of compressible flows in nozzles, a behavioral classification of non-adiabatic compressible flows is proposed through the definition of critical pressure ratios. It is demonstrated that a critical pressure ratio distinguishes subsonic and supersonic outlet regimes and that there cannot be a steady shock-wave in such configuration. The behavior of this critical pressure ratio is studied for limit cases of heat-flux, delineating physical boundaries. An abacus is also proposed for a given couple of heating and cooling powers, as both values are needed to characterize the flow. Results are studied for parameters such as pressure ratio and outlet heat power. A short validation of a numerical simulation tool is provided, yielding excellent results and very small relative errors.
\end{abstract}
    \section{Introduction}
\label{sec:intro}

This paper proposes analytical or semi-analytical solutions of one-dimensional, single phase, steady compressible flows in a non-adiabatic duct. We propose the following simple question: \textit{What are the possible solutions for a compressible flow in a duct of length $L$ receiving a certain amount of power $Q_{out}$ ?}

This problem has had quite established solutions, but they may appear too restrictive for actual applications. Simple studies can be found in \cite{hicks1945} or in \cite{shapiro1953dynamics} : they exhibit analytical solutions for so-called "$T_0$-change". Indeed, the heat power applied to the flow will increase (heating) or decrease (cooling) its stagnation temperature, denoted by $T_0$. The previously mentioned analyses have shown a specific solution called "choked solution", which corresponds to a Mach number equal to unity, obtained for a maximum heat. Author's analysis by \cite{shapiro1953dynamics} (p.203) suggests that if the heat is larger than this authorized value, the change of flow conditions is done upstream of the duct.

To obtain those solutions, the flow conditions (such as velocity, pressure, temperature) must be specified at the inlet. However, in subsonic regime, it is not possible to keep the quantities fixed as waves will travel up and down the duct, which will change the inlet state ; only the inlet stagnation state conditions are constant. Therefore, in realistic context, the previously proposed analysis cannot be applied to recover solutions of actual flows (without questioning the rigor of the analysis).

In the following study, upstream variations are taken into account to determine solutions, by connecting the inlet of the duct to a tank with fixed stagnation conditions.
We examine the case where the heat is not uniformly distributed along the duct, but instead obeys a position-dependent function. In doing so, the analysis can be reduced to a solely heated or cooled flow. The solutions are extended to the "Stiffened-Gas" equation of state, allowing to consider both liquid flows and ideal gases. 

Beyond furthering the knowledge on compressible flows, these analytical solutions will help to give preliminary insights on the behavior of a more complex system than what has been developed thus far. Indeed, compressible fluids with high-speed motion have been studied for military, aeronautical and spatial purposes \cite{hicks1945}: as example, cryogenic fluids are used into spatial systems, for propulsion purposes, but also to cool engine and nozzles walls that are subjected to several MW/$\text{m}^2$ of thermal heat fluxes.

Realistic and quantitative numerical methods and models have therefore been developed to perform more complex simulations of non-adiabatic flows \cite{bermudez2016}, \cite{schmidmayer2018}. However, those simulations often remain confidential given the fields of research, or too restrictive (limited to a specific geometry, element, regime). Hence, the solutions proposed in this study could be used as future reference solutions for the validation of numerical tools (examples will be presented in Section \ref{sec:results}, applied to ECOGEN open-source code by \cite{schmidmayer2018}).

In this paper, we first recall the theoretical elements for flows moving in non-adiabatic ducts. Basic local relations are exhibited in the extended case of "Stiffened-Gas" equation of state. In the second part, a more global solution is developed and provides a full description of the flow behavior in a heated/cooled duct at steady state. The third part is dedicated to the study of some limit cases that leads to the plotting of abacus useful to discriminate subsonic and supersonic flow regimes. In the final part, typical results are presented and analyzed for flows in heated and cooled ducts. Finally, the usefulness of these reference solutions is proved by validating a numerical tool on two practical cases (for subsonic and supersonic behaviors).
    
\section{Model for non-adiabatic single-phase flows}
\label{sec:model}

\subsection{Unsteady model}

Although unsteady solutions are not addressed in this study, the governing equations are presented down below as they lay the ground for the problem that will be dealt with in steady conditions. For non-adiabatic compressible inviscid single phase flows, the fluid state at each point is governed by the Euler equations, including heat source term:

\begin{equation}
\label{eqn:euler}
    \begin{cases}
        \frac{\partial \rho}{\partial t} + \text{div}(\rho \textbf{u}) = 0 \\
        \frac{\partial \rho \textbf{u}}{\partial t} + \text{div}(\rho \textbf{u} \textbf{u} + P\textbf{I}) = 0 \\
        \frac{\partial \rho E}{\partial t} + \text{div}((\rho E + P)\textbf{u}) = \delta \dot q
    \end{cases}
\end{equation}

where $\rho$ represents the density, \textbf{u} the velocity vector, $P$ the pressure and $E=e+\frac{1}{2}\textbf{u}^2$ the total specific energy, with $e$ the internal specific energy. $\delta \dot q$ represents the received (or lost) thermal power per volume unit ($\text{W/m}^3$) of fluid.

The entropy equation is:
\begin{equation}
    \label{eqn:entropy}
    \frac{\partial\rho s}{\partial t} + \text{div}(\rho s \textbf{u}) = \frac{\delta \dot q}{T}
\end{equation}
where $T$ represents the temperature of the fluid and $s$ the specific entropy.

System (\ref{eqn:euler}) is closed by any convex equation of state (EOS) $e=e(P,v)$. In the present work, the "Stiffened Gas" (SG) EOS is used:

\begin{equation}
    \label{eqn:energy}
    e(P,v) = \frac{P + \gamma P_{\infty}}{\gamma -1}v+e_{ref}
\end{equation}

This EOS allows to describe the thermodynamic properties of both liquids and gases, as it takes into account both the repulsive effects and the attractive molecular properties that binds the matter together. The EOS parameters $\gamma$, $P_{\infty}$ and $e_{ref}$ are obtained from reference thermodynamic curves, characteristics of the material and transformation under study. See \cite{lemetayer2004} for details. When $P_\infty$ is null, the SG EOS can be simplified to the Ideal Gas (IG) EOS.

\subsection{Steady-state analysis}

\subsubsection{General problem}

Let us consider a one-dimensional flow in a rectilinear pipe, defined by a length $L$, a constant cross-section $S$ and a circumference $C$, receiving or losing a quantity of heat flux $\varphi$ over its external surface $S_{ext}=C \times L$.
Its inlet state quantities are denoted by subscript $(\cdot)_{in}$.

\begin{figure*}
\centering
    \includegraphics[width=0.6\textwidth]{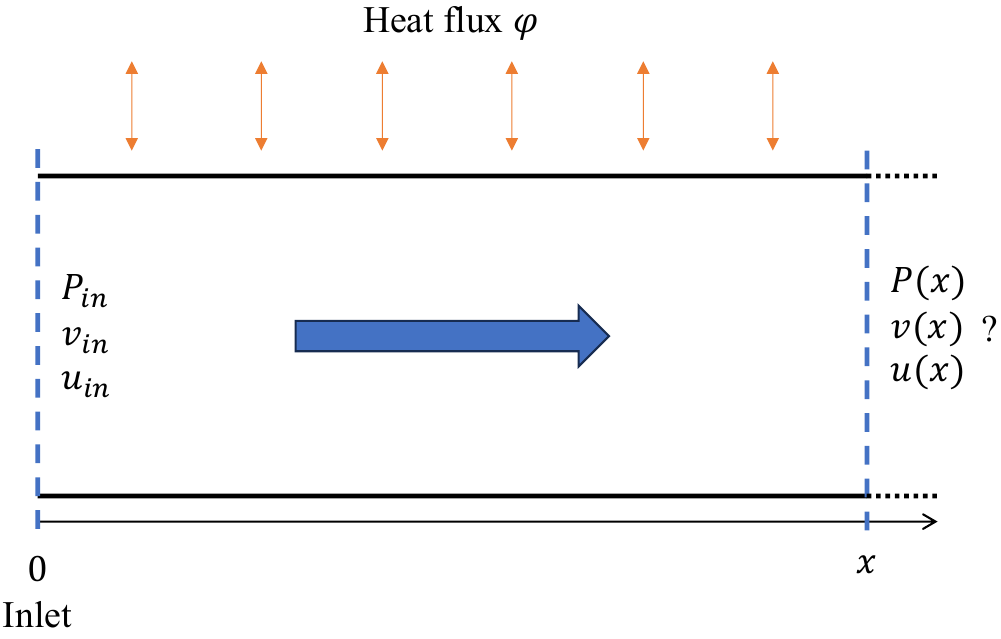}
    \captionof{figure}{Flow in a duct receiving or losing a heat flux}
    \label{fig:conduite_simple}
\end{figure*}

We will consider two cases to obtain mono-dimensional steady-state solutions of system~(\ref{eqn:euler}). The first one (one can call the \textit{local solution}) is presented in \cite{shapiro1953dynamics}. The inlet quantities are imposed. Hence, we may then ask ourselves what the flow solution is at a distance $x$ from the inlet, denoted by subscript $(\cdot)_{x}$, as the duct has received a certain amount of heat flux $\varphi$. The configuration is depicted in figure~\ref{fig:conduite_simple}.

The second case is related to a more global solution, which is supposed to be more realistic. At the inlet, only the stagnation conditions are prescribed. At the outlet, denoted by subscript $(\cdot)_{out}$, the pressure is prescribed: in the case of the subsonic flow, it is equal to the external one. In the supersonic case, no quantities can be imposed. This configuration will be described in depth later on.


\subsubsection{Local solution}

The following relations are \textit{local} because they can be used whatever the location of $x$. 
Corresponding to the configuration presented in figure~\ref{fig:conduite_simple}, integration of system (\ref{eqn:euler}) at steady-state between the two points $(\cdot)_{in}$ and $(\cdot)_x$ along axis $x$, results in the following system:
\begin{equation}
    \label{eqn:mass}
    \rho_{in}u_{in} = \rho_{x} u_{x} = \dot m_{s} 
\end{equation}

\begin{equation}
    \label{eqn:qdm}
    \rho_{in}u_{in}^2 + P_{in} = \rho_{x}u_{x}^2 + P_{x}
\end{equation}

\begin{equation}
    \label{eqn:nrj}
    \dot m_{s}(H_{x} - H_{in}) = \dot q_{s,x}
\end{equation}

where $\dot m_{s} = \rho u = \dot m/S$ is the mass flow rate per unit area of cross-section, $\dot q_s=Q/S$ is the power received per unit area of cross-section and $H=h+\frac{1}{2} u^2$ is the total specific enthalpy.

Combining equations (\ref{eqn:mass}) and (\ref{eqn:qdm}) leads to the equation of the \textit{Rayleigh line} in the $(P,v)$ plane:
\begin{equation}
    \label{eqn:rayleigh}
    P_{x} = \dot m_{s}^2(v_{in} - v_{x}) + P_{in}
\end{equation}

For a given steady flow, $\dot m_{s}$ is constant. Thus for a known inlet state, $-\dot m_{s}^2$ is the slope of its  \textit{Rayleigh line} in the $(P,v)$ plane.

The definition of internal specific enthalpy $h=e+Pv$ combined to equations (\ref{eqn:mass}) and (\ref{eqn:qdm}) leads to equation (\ref{eqn:nrj}) being rewritten as:

\begin{equation}
    \label{eqn:demohugo}
    e_x(P_x, v_x) - e_{in}(P_{in}, v_{in}) + \frac{P_{in} + P_{x}}{2}(v_x - v_{in}) = \frac{\dot q_{s,x}}{\dot m_s} 
\end{equation}

The thermodynamic closure is achieved thanks to a convex EOS $e(P,v)$ (\ref{eqn:energy}). Using the SG EOS, an explicit relation between the pressures and specific volumes between states $(\cdot)_{in}$ and $(\cdot)_x$ is obtained from (\ref{eqn:demohugo}):

\begin{equation}
    \label{eqn:crussard}
    P_x =(P_{in} + P_{\infty}) \frac{ (\gamma+1) v_{in} - (\gamma-1)v_x }{ (\gamma+1) v_{x} - (\gamma-1)v_{in} } + \frac{ 2(\gamma-1) }{ (\gamma+1) v_{x} - (\gamma-1) v_{in} } \frac{\dot q_{s,x}}{\dot m_s} - P_{\infty}
\end{equation}

This relation describes the \textit{Crussard curve} in the $(P,v)$ plane. It can be rewritten to express the specific volume at point $(\cdot)_x$:

\begin{equation}
\label{eqn:v_crussard}
    v_{x}(P_{in}) = \frac{ (\gamma+1)(P_{in} + P_{\infty}) + (\gamma-1)(P_{x} + P_{\infty}) }{ (\gamma+1)(P_{x} + P_{\infty}) + (\gamma-1)(P_{in} + P_{\infty}) } v_{in} + \frac{ 2(\gamma-1)  }{ (\gamma+1)(P_{x} + P_{\infty}) + (\gamma-1)(P_{in} + P_{\infty}) } \frac{\dot q_{s,x}}{\dot m_s}
\end{equation}

The \textit{Hugoniot curve} is obtained in the limit case for an adiabatic transformation, where $\dot q_{s,x}=0$:
\begin{equation}
    \label{eqn:hugoniot}
    P_x = (P_{in} + P_{\infty}) \frac{ (\gamma+1) v_{in} - (\gamma-1)v_x }{ (\gamma+1) v_{x} - (\gamma-1) v_{in} } - P_{\infty}
\end{equation}

\begin{figure*}
    \centering
    \includegraphics[width=0.85\textwidth]{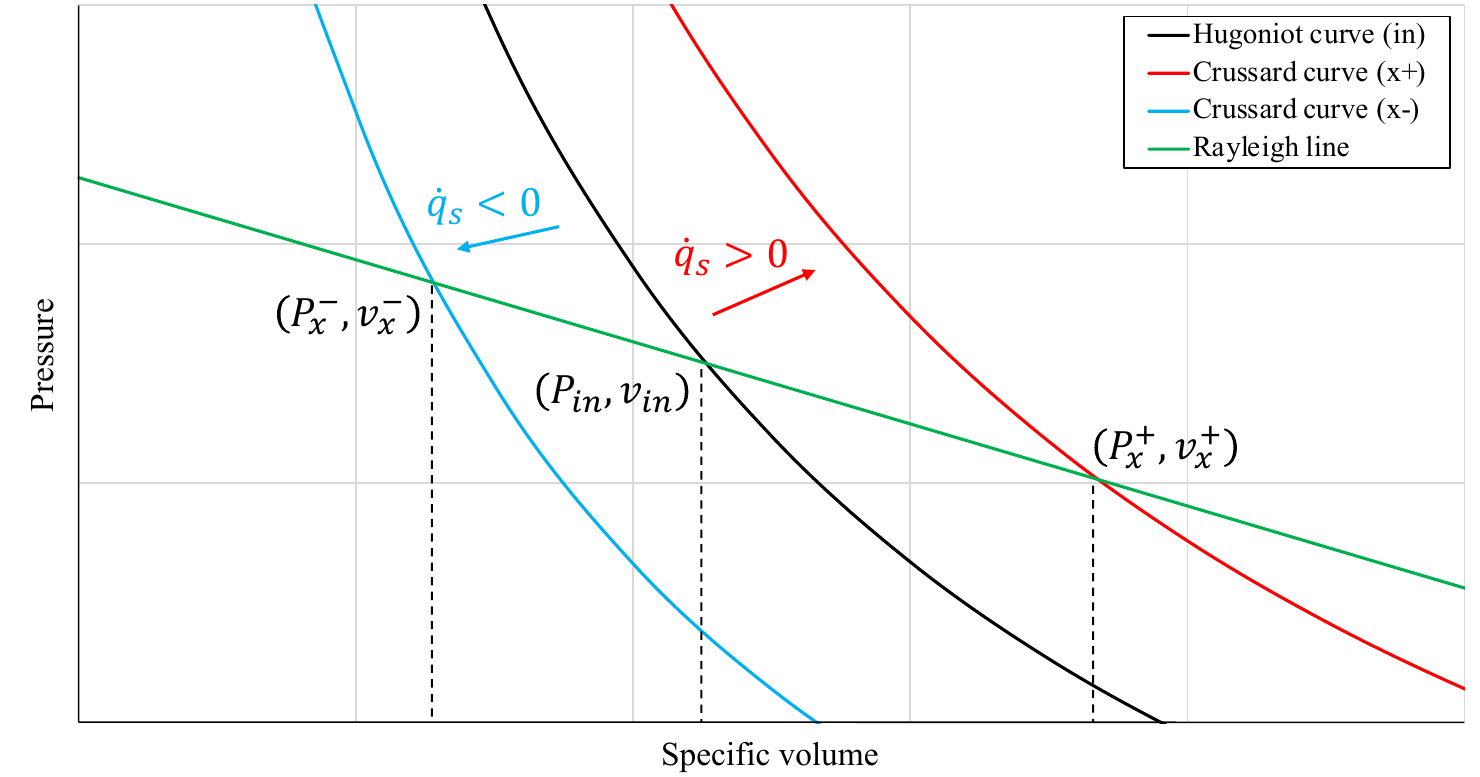}
    \caption{Thermodynamic path for a fluid flowing in a heated/cooled duct in the $(P,v)$ plane. Starting from state $(\cdot)_{in}$, the fluid can be heated to reach $(\cdot)_{x}^{+}$ state or cooled down to reach $(\cdot)_{x}^{-}$ state.}
    \label{fig:PV}
\end{figure*}

The thermodynamic path of fluid between states $(\cdot)_{in}$ and $(\cdot)_x$ is represented in figure~\ref{fig:PV}. Each state is constrained by both the \textit{Rayleigh line} and its respective \textit{Crussard curve}. This figure clearly shows that for the imposed inlet flow conditions, only one solution is reachable at the location $x$ according the sign of the power received, given by the intersection of the \textit{Rayleigh line} and the corresponding \textit{Crussard curve}.

\subsubsection{Local solution in Mach number}

Using the SG EOS, Euler equations between $(\cdot)_{in}$ and $(\cdot)_{x}$ can be also rewritten in terms of Mach number. Indeed, analysis of the solution in Mach number allows to obtain information about the global  dynamics of the flow. Moreover, this number is quite pertinent and usual for the analysis of compressible flows.

The Mach number is defined by $M=u/c$, with $c$ the speed of sound defined as $c^2 = \frac {\partial P}{\partial \rho}\bigr)_s$. In the case of the SG EOS, the speed of sound reads:

\begin{equation}
    \label{eqn:celerity}
    c = \sqrt{ \frac{\gamma (P + P_{\infty})}{\rho}  }
\end{equation}

Equation (\ref{eqn:qdm}) is thus expressed as a ratio of pressure:
\begin{equation}
    \label{eqn:PMach}
    \frac{P_{in} + P_{\infty}}{P_{x} + P_{\infty}} = \frac{1 + \gamma M_{x}^2}{1 + \gamma M_{in}^2}
\end{equation}

The equation (\ref{eqn:PMach}) coupled to (\ref{eqn:mass}) allows to express the ratio of temperature:
\begin{equation}
    \label{eqn:TMach}
    \frac{T_{in}}{T_{x}} = \frac{M_{in}^2 (1 + \gamma M_{x}^2)^2}{M_{x}^2 (1 + \gamma M_{in}^2)^2}
\end{equation}

Taking into account for heat variation, the so-called $T_0$-change (change in stagnation temperature) between the inlet $(\cdot)_{in}$ and point $(\cdot)_x$ is a direct measure of the amount of heat input. The energy equation (\ref{eqn:nrj}) becomes:
\begin{equation}
    \label{eqn:stagnantenth}
    h_{0,x} - h_{0,in} = C_p(T_{0,x} - T_{0,in}) = \frac{\dot q_{s,x}}{\dot m_s}
\end{equation}

The inlet stagnation state, which is retrieved by bringing the flow at rest through an adiabatic process 
can be expressed as a function of Mach number. This yields the following isentropic relations:

\begin{equation}
    \label{eqn:stagnantpressure}
    \frac{P_{0,in} + P_{\infty}}{P_{in} + P_{\infty}} = \left( 1 + \frac{\gamma-1}{2} M_{in}^2 \right)^{\frac{\gamma}{\gamma-1}}
\end{equation}

\begin{equation}
    \label{eqn:stagnanttemp}
    \frac{T_{0,in}}{T_{in}} = 1 + \frac{\gamma-1}{2} M_{in}^2
\end{equation}

The combination of equations (\ref{eqn:TMach}) and (\ref{eqn:stagnanttemp}) allows to find a relation between the Mach number at the inlet $(\cdot)_{in}$ and at point $x$, through a ratio of stagnation temperatures:
\begin{equation}
    \label{eqn:T0_ratio}
    \frac{T_{0,in}}{T_{0,x}} = \frac{ M_{in}^2 (1+\gamma M_x^2)^2 (1 + \frac{\gamma-1}{2} M_{in}^2) }{ M_x^2 (1+\gamma M_{in}^2)^2 (1 + \frac{\gamma-1}{2} M_x^2) }
\end{equation}

\begin{figure*}
    \centering  
    \includegraphics[width=0.85\textwidth]{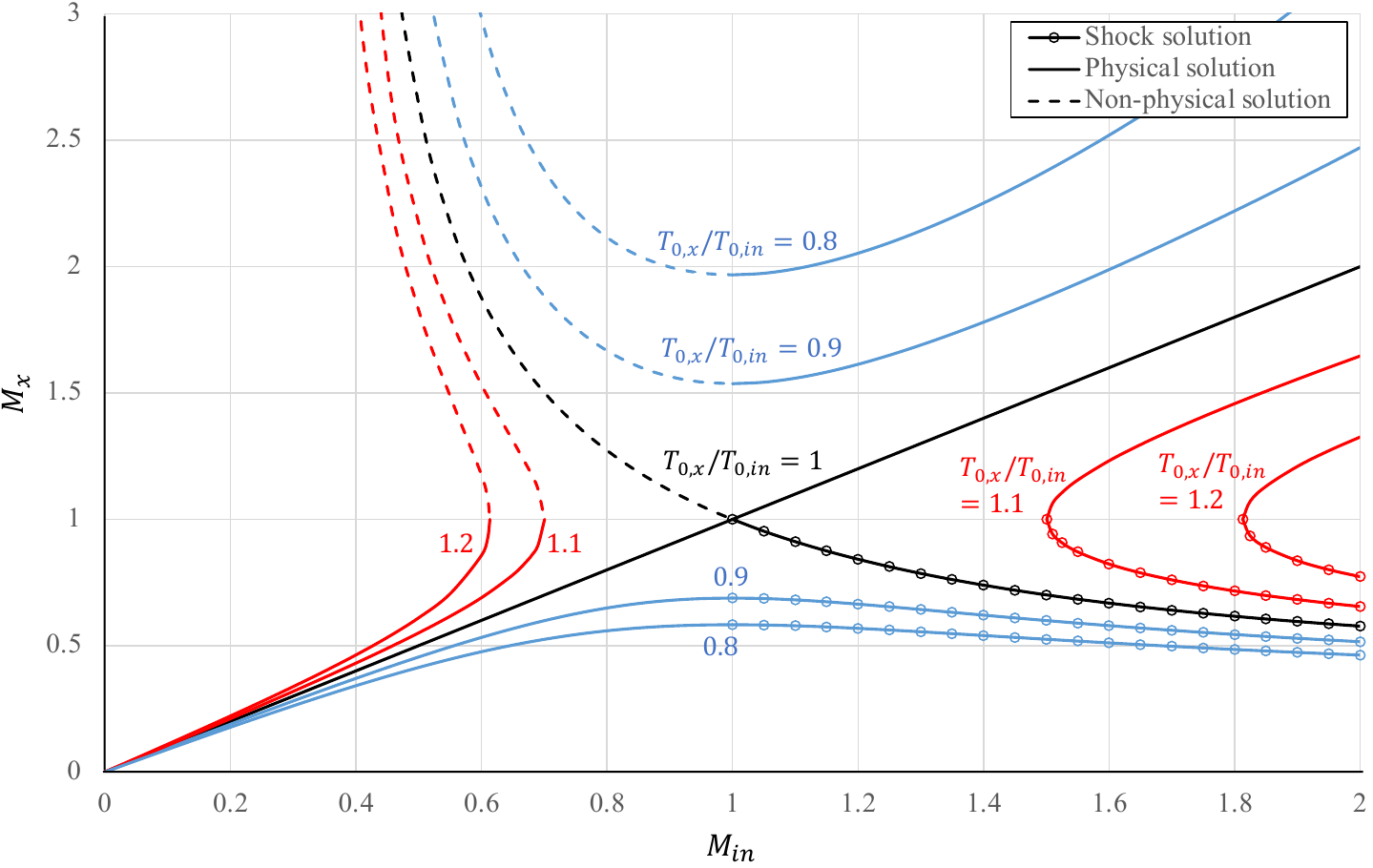}
    \caption{\textit{Local solution} for $M_x$ versus $M_{in}$: when the inlet is subsonic ($M_{in} < 1$), the only physical solution is a subsonic outlet ($M_x < 1$). When the inlet is supersonic ($M_{in} > 1$), the solution can either be supersonic ($M_x > 1$), or subsonic through a shock ($M_x < 1$).}
    \label{fig:T0ratio_big}
   
\end{figure*}

The analysis of these solutions is detailed in literature \cite{shapiro1953dynamics} and show how complex a compressible flow can be. Important results are summarized in figure~\ref{fig:T0ratio_big}, which illustrates the behavior of the Mach number at point $x$ $M_{x}$ versus the inlet Mach number $M_{in}$, for fixed values of ratios of stagnation temperatures $T_{0,x}/T_{0,in}$. 
Obviously, $T_{0,x}/T_{0,in}=1$ represents adiabatic solutions, where either no change of flow regime happens or a shock wave is a possible solution.
For $T_{0,x}/T_{0,in}\neq1$, we are either in the case of heating ($T_{0,x}/T_{0,in}>1$) or cooling ($T_{0,x}/T_{0,in}<1$). On these branches, there are two possible values of $M_{x}$ for one value $M_{in}$. When the inlet is supersonic, the outlet can also be directly supersonic, or subsonic through a shock wave. However, for initially subsonic inlet cases, the only possible physical outcome is also a subsonic flow \cite{shapiro1953dynamics}.

\subsubsection{Limitations of \textit{local solutions}}

The previous configuration that corresponds to the \textit{local solution} is too far of a realistic system and is not self-sufficient to describe the full reality of the flow. Indeed, in a subsonic flow, the inlet continuously re-adapts to the external conditions (such as outlet pressure and heat flux) and cannot be fixed. 
It means that one cannot impose the inlet conditions and then apply a power heat to calculate the solution at point $x$: doing so, travelling acoustics waves will change the inlet conditions. Hence, the conditions at the inlet of the duct are no longer known, meaning that the flow along the duct cannot be determined.
Notice that for a supersonic flow, the prescription of the inlet state is direct as acoustic perturbations cannot go back upstream : the local solution is then valid.
Besides, in actual system, the stagnation conditions are generally known and the flow is then governed by the external pressure.

\begin{figure*}
    \centering
    \includegraphics[width=0.6\textwidth]{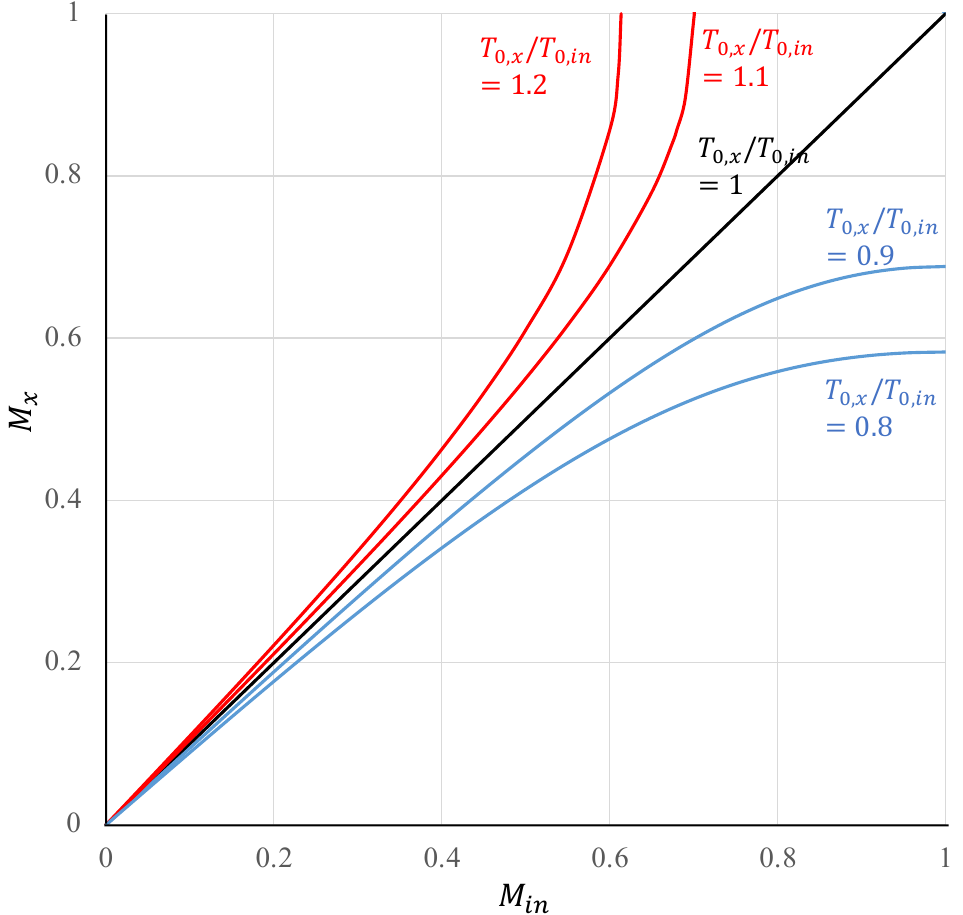}
    \caption{\textit{Local solution} for $M_x$ versus $M_{in}$: focus on the subsonic inlet area ($M_{in}<1$), which can only lead to a subsonic or sonic outlet ($M_x \leq 1$), whatever the heat-flux applied.}
    \label{fig:T0ratio_focus}
\end{figure*}

So, we propose a study that is closer to this configuration, by considering imposed tank conditions instead of inlet conditions. In the tank, the velocity is equal to zero while the pressure $P_0$ and temperature $T_0$ are given data.
It also means that at the flow regime in the section connected to the tank is necessarily subsonic, that is to say $M_{in} <1 $. The possible solutions for Mach number are then depicted in figure~\ref{fig:T0ratio_focus}, which is a restriction of the solutions plotted in figure~\ref{fig:T0ratio_big}: on one hand, locally heating increases the flow velocity (curve slope over 1) up until the sonic point ; on the other hand, cooling decreases the flow velocity (curve slope under 1).

In the following analysis, a complete description of the flow is proposed. Stemming from the previous part, steady relations will be established in two parts: the first set of relations are the local equations adapted in the case of interest to obtain the flow solution along the duct. 
The second set of relations is between the inlet of the duct and the tank at the stagnation conditions; notice that the inlet flow conditions are unknown in this configuration. Coupling both systems will allow to define the flow characteristics along the duct, for given stagnation conditions, an outlet pressure and an applied heat flux function.

    \section{Global solution in a duct}
\label{sec:glob_sol}

\subsection{Observations}


Before delving into a mathematical description of the considered flows, we present thereafter some typical qualitative results of flows in a duct. Some flow behaviors are easily deduced from isentropic or adiabatic flow analysis, or from the basic results \cite{shapiro1953dynamics} depicted in figure~\ref{fig:T0ratio_big}, while the more complex ones are presented in light of the subsequent analysis. For now, the aim is simply to highlight the key parameters of the flow, in order to familiarize the reader with its behavior in this new configuration.

We describe a flow in a duct subjected to a total amount of power $Q=\sum Q_i$, where $Q_i$ is a local applied amount of heating/cooling power, which can be positive or negative. This study is inspired by the study of adiabatic flows in nozzles: in such flows, the area of the cross section of the nozzle is of prime importance, through the ratio of the area over the critical area. In the present work, the local amount of heat will play this role. Nevertheless, the characteristic of the configurations (area for nozzle, heating for the duct) is not sufficient to determine flows conditions: they are driven by the external pressure $P_{out}$ and more accurately by the pressure ratio defined as $\Pi=P_{out}/{P_0}$. Because the tank conditions are given fixed data, the only way to create a flow through the nozzle or the duct is to vary the outlet pressure $P_{out}$, like one would do opening or closing a valve of a vacuum pump. To generate a flow from the tank to the outlet of the duct, the pressure is chosen accordingly and $\Pi$ is necessarily less than $1$.

\subsubsection{Isentropic flow}
    When the duct is not subjected to any heat flux, the flow is defined by only one state (constant along the duct) which can be recovered through equations~(\ref{eqn:stagnantpressure}) and (\ref{eqn:stagnanttemp}): its pressure is equal to the external pressure $P_{out}$.
    
    Decreasing $\Pi = P_{out}/P_0$ leads to an increase of the flow velocity. As the inlet of the duct is connected to the tank, isentropic relations govern the flow between both, hence why the inlet state can only be subsonic. Therefore, the velocity increases to reach the sonic value, as the pressure ratio is decreased:  
   then, a sonic choking of the flow would occur at the outlet and the downstream pressure would no longer be able to affect the flow inside the duct, remaining at the sonic state. 
    
\subsubsection{Heated flow}
    Now the duct is subjected to a fixed heating power $Q_{heat}>0$. Because of the heat exchange, the flow cannot be constant in the duct: indeed, the amount of heat received at a location $x$ is less than the heat received at the location $x+\text{d}x$. Therefore, the inlet will adapt to the outlet pressure in the way described by figure~\ref{fig:T0ratio_focus}: the inlet Mach number $M_{in}$ (always subsonic) is necessary less than $M_{x}$ at any position $x$ in the duct. So, for a heating power, the flow will accelerate all along the duct.
    
\begin{figure*}
\centering
    \begin{subfigure}[b]{0.49\textwidth}
        \includegraphics[width=\textwidth]{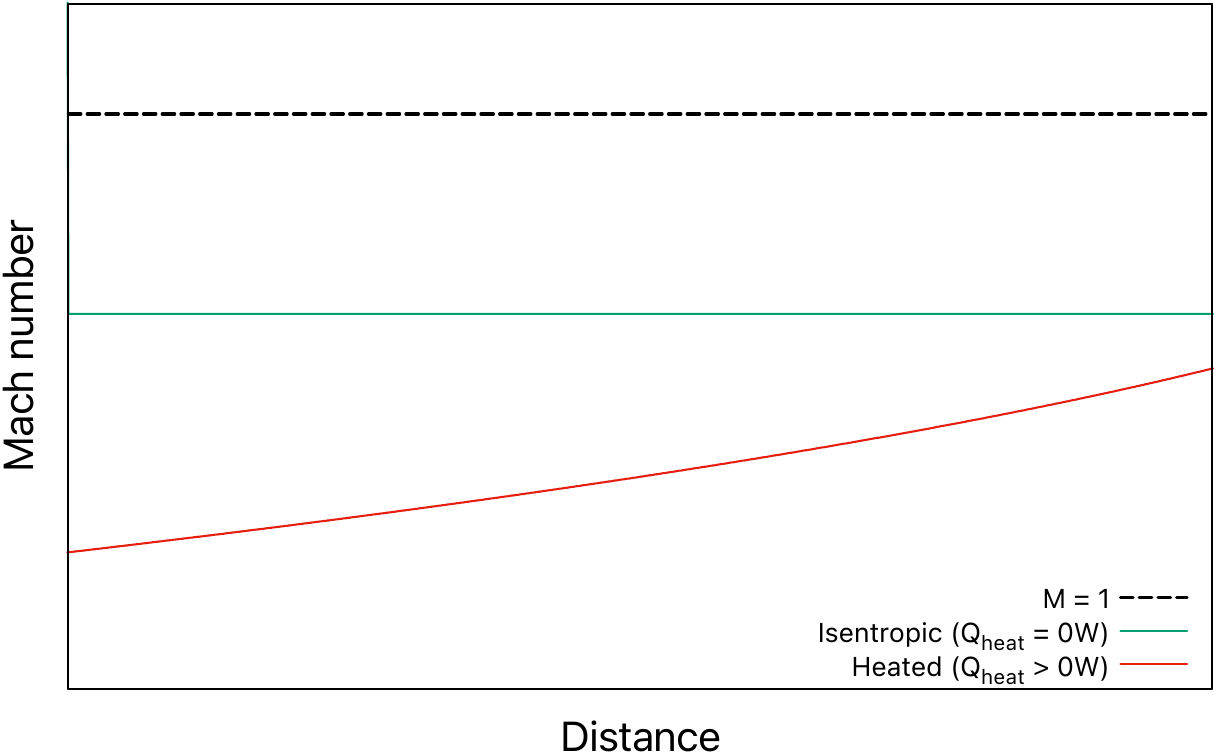}
        \label{fig:obs_Mach}
    \end{subfigure}
    \begin{subfigure}[b]{0.49\textwidth}
        \includegraphics[width=\linewidth]{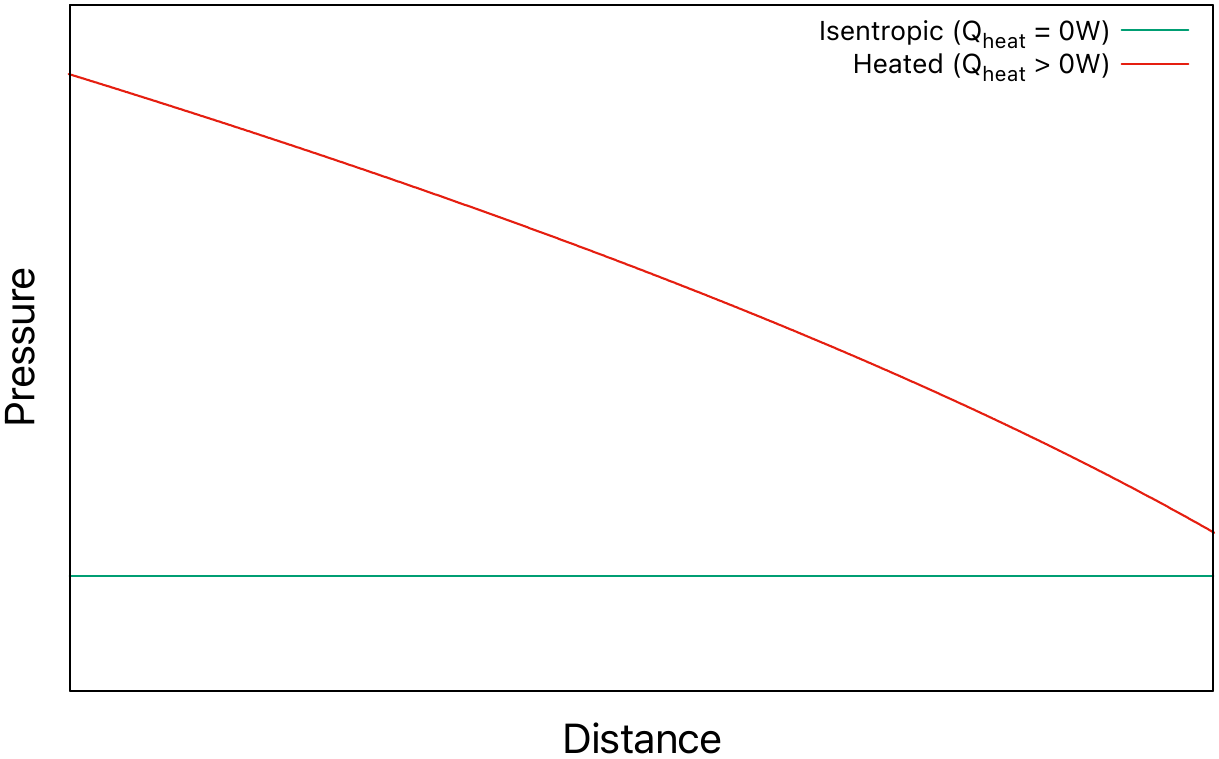}
        \label{fig:obs_P}
    \end{subfigure}
    \caption{Qualitative behavior of a flow within a duct connected to a tank and an outlet pressure with a subsonic inlet, for different heating powers: Mach number (left) and pressure (right).}
    \label{fig:observations}
\end{figure*}

    Moreover, the Mach number cannot become supersonic to maintain the physicality of the system, as figure~\ref{fig:T0ratio_focus} shows that with a subsonic inlet, the outlet must also be subsonic. Considering the pressure, its evolution along the duct is on the opposite of the Mach number, meaning that a decrease will be observed between the inlet and the outlet cross sections. The behavior of these quantities are illustrated in figure~\ref{fig:observations}, where Mach and pressure profile are plotted for a given heating powers.  

\subsubsection{Cooled flow}
    The flow is now cooled down by a fixed cooling power $Q_{cool}<0$. The pressure and Mach profiles evolve in the opposite way to the heated case: pressure is increasing between inlet and exit section, while the flow is decelerated.
    
    Hence, when $\Pi$ is decreased, 
    the inlet pressure decreases and the inlet Mach number rises to become supersonic: this is not part of our study as it will not respect the hypothesis of a subsonic inlet. 

    Obviously, the upper value of the ratio $\Pi$ is bounded by unity ; this short analysis indicates that some lower limits may exist and could be exhibited by deeper calculations.

\subsubsection{Heated and cooled flow}
    In this work and in order to complete our observations, we combine both configurations: after the heated subsection receiving $Q_{heat}$, a subsection subjected to a fixed cooling power $Q_{cool}$ is added. For a rather high pressure ratio $\Pi$ (but still bounded by 1), the flow remains subsonic: the heated part accelerates, whereas the cooled part decelerates.
    When the outlet pressure is decreased, the flow overall accelerates as the end part of the heated subsection is brought closer to the sonic point. 
    
    Once $\Pi$ gets below a certain value, the flow reaches the sonic point at the end of the heated part. Then, the flow in the cooled part can become subsonic or supersonic because both solutions are possible. It suggests that this bifurcation is governed by specific pressure ratios called \textit{critical pressure ratios}. 
    
\subsubsection{Cooled and heated flow}
    This configuration has no theoretical interest and the whole solution is provided by the analysis of figure~\ref{fig:T0ratio_focus} that shows the flow will decelerate and accelerate while the outlet section will remain subsonic.

This short description of flow receiving a power shows the outlines of the study that will be presented hereafter: the solutions for either heated or cooled flows are well documented in \cite{shapiro1953dynamics}. Cooled and heated duct flows solutions are easily deduced from the previous observations. On the contrary, the solutions for heated and cooled duct flows seem to exhibit non trivial solutions. 
The next part of the study will focus on this configuration.

\subsection{System description}

We propose to investigate an uneven heating distribution on the duct, which consists in an extension of the work done in \cite{fathalli2018}. This will allow to describe as many cases as possible, as we have seen that it is now possible to reach either a subsonic outlet or a supersonic outlet with a subsonic inlet in such configuration.

\begin{figure*}
    \centering
    \includegraphics[width=0.85\textwidth]{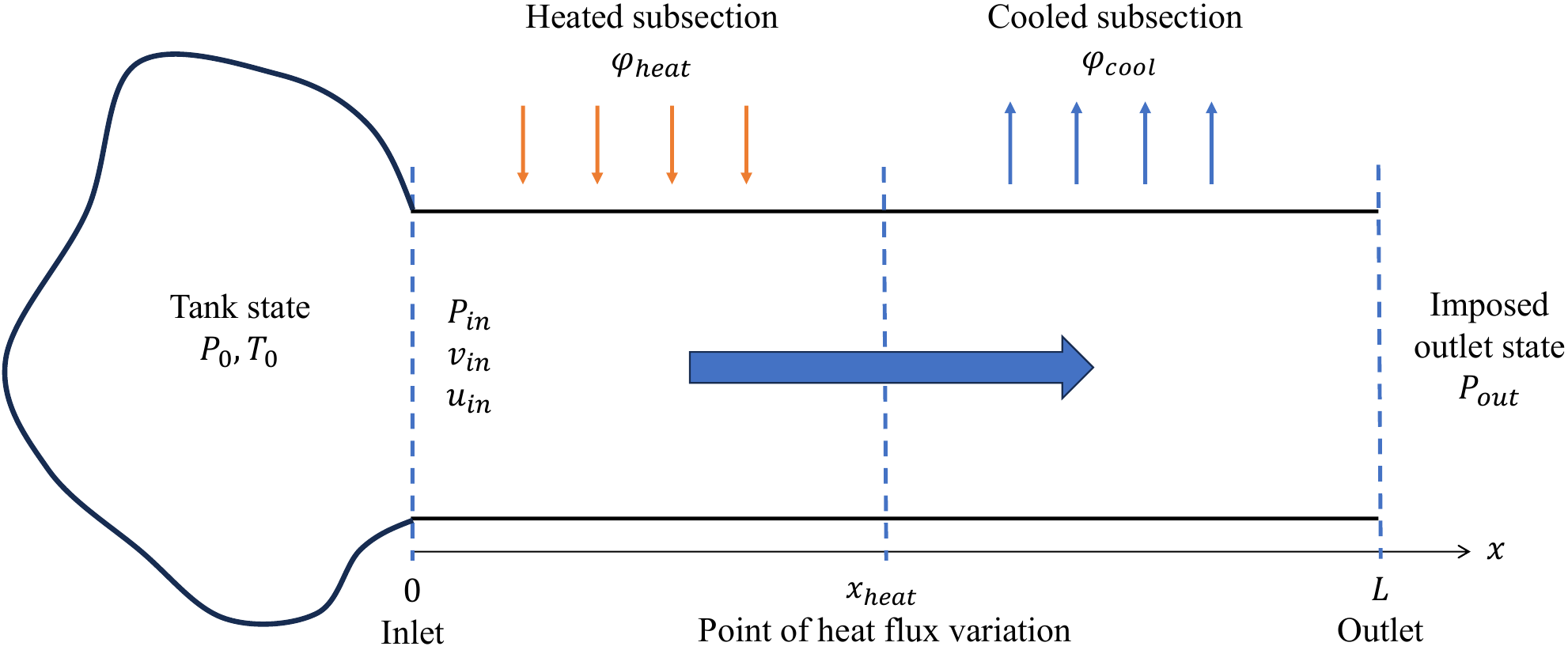}
    \caption{Tank flowing in a heated and cooled pipe}
    \label{fig:system}
\end{figure*}

To study the flow, we consider a duct connected to a tank (denoted by subscript $(\cdot)_0$) whose state is supposed known (pressure and temperature), and an outlet pressure imposed by external conditions. The whole duct is governed by non-adiabatic relationships as it receives or loses a certain amount of heat flux, whereas the tank and inlet states are coupled by isentropic relationships. The system is depicted in figure~\ref{fig:system}.

\subsubsection{Duct relations}

The duct under study is divided in two sub-sections, separated by the point of variation of flux distribution (defining the subscript $(\cdot)_{heat}$): a heated subsection between $x=0$ and $x=x_{heat}$, followed by a cooled subsection between $x=x_{heat}$ and $x=L$. The local relations exposed previously are suitable to describe the behavior of the flow in the duct.

Along axis $x$, a function of heat flux $\varphi(x)$ $(W.m^{-2})$ is distributed on the external surface of the duct:
$$
\varphi(x) = \left\{
    \begin{array}{ll}
        \varphi_{heat} > 0 & \mbox{if } x \leq x_{heat} \\
        \varphi_{cool} < 0 & \mbox{otherwise}
    \end{array}
\right.
$$

\begin{figure*}
\centering
    \begin{subfigure}[b]{0.49\textwidth}
        \includegraphics[width=\textwidth]{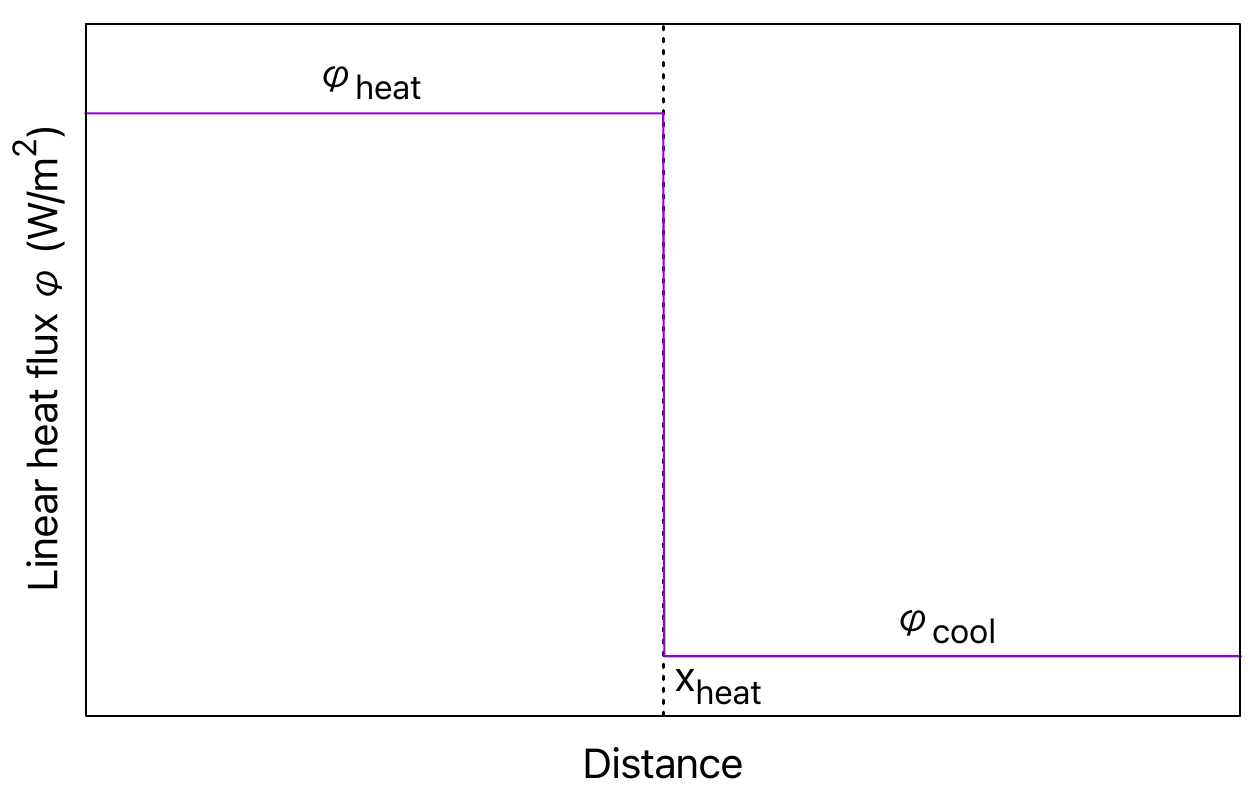}
        \label{fig:phi_x}
    \end{subfigure}
    \begin{subfigure}[b]{0.49\textwidth}
        \includegraphics[width=\textwidth]{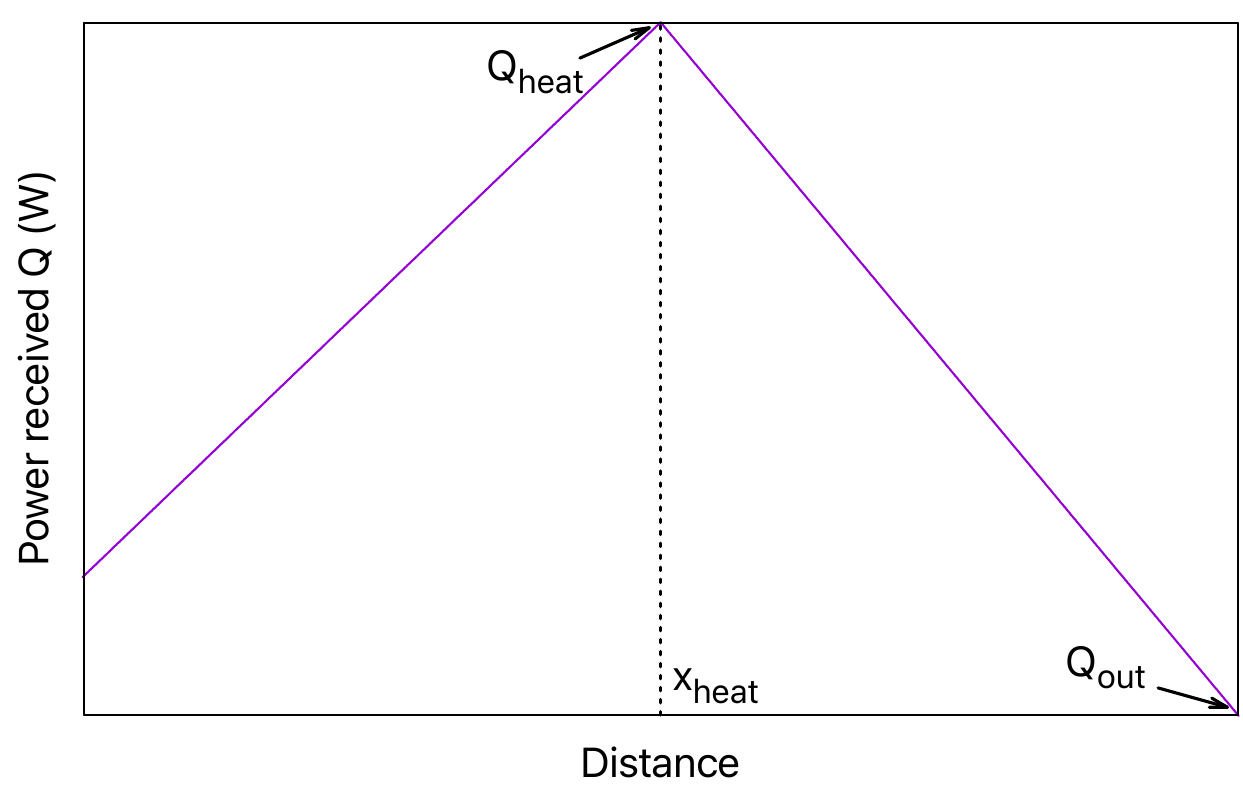}
        \label{fig:Q_x}
    \end{subfigure}
    \caption{Relationship between heat flux variables: $\varphi(x)$ (left) and $Q(x)$ (right)}
    \label{fig:heatfluxes}
\end{figure*}

The power $Q(x) = \int_0^x \varphi(\eta) \, C \, \mathrm{d}\eta$ $(W)$ applied to the flowing fluid is determined by the distribution of $\varphi(x)$ (figure~\ref{fig:heatfluxes}):
$$
Q(x) = \left\{
    \begin{array}{ll}
        C \, x \, \varphi_{heat} & \mbox{if } x \leq x_{heat} \\
        C \left[ x_{heat} \, \varphi_{heat} + (x-x_{heat}) \, \varphi_{cool} \right] & \mbox{otherwise}
    \end{array}
\right.
$$

The power received by the flow on the entire heated subsection is $Q_{heat} = C \, x_{heat} \, \varphi_{heat}$, the power lost by the flow on the entire cooled subsection is $Q_{cool} = C \, (L-x_{heat}) \, \varphi_{cool}$ and the power received by the flow on the entire duct is $Q_{out} = Q_{heat} + Q_{cool}$. If $Q_{out}>0$, the duct has received more heat than it has lost. Conversely, if $Q_{out}<0$, the duct has lost more heat than it has received. Finally, if $Q_{out}=0$, we can consider the outlet flow adiabatic.

The power received per unit area of cross-section is defined as : $\dot q_{s}(x) = Q(x)/S$ ($W.m^{-2}$). As we apply a function of heat flux $\dot q_s(x) \varpropto Q(x)$ on each point $x \equiv (\cdot)_x$ of the duct, equations  (\ref{eqn:rayleigh}) and (\ref{eqn:crussard}) govern the flow between the inlet and any point of the duct. The applied heat flux can be directly correlated to the notion of $T_0$-change through equation~(\ref{eqn:stagnantenth}):

\begin{equation}
\label{eqn:T0x/T0in}
    \frac{T_{0,x}}{T_{0,in}} = \frac{\dot q_{s,x}}{\dot m_s C_p T_{0,in}} + 1
\end{equation}

\subsubsection{Tank-inlet relations}

The closure of the system is thus done by connecting the inlet of the duct to a tank. The inlet stagnation state is recovered by slowing down the flow to rest through an isentropic process and is equivalent to the tank state. This means that $P_{0,in} \equiv P_0$ and $T_{0,in} \equiv T_0$.

On the \textit{tank-inlet} side, the only available relations are: $H_{in} = H_0$, $s_{in} = s_0$ where $H$ is the total specific enthalpy, and $s$ is the specific entropy. Even if the supersonic solution is mathematically acceptable, only the subsonic solution is sought to be consistent with an actual tank. 
The first condition $H_{in} = H_0$ allows us to establish a relation linking the mass flow rate and the inlet pressure:
\begin{equation}
    \label{eqn:massflowrateinlet}
    \dot m_s(P_{in}) = \frac{\sqrt{2(h_0 - h_{in}^{is})}}{v_{in}^{is}} 
\end{equation}

$v_{in}^{is}(P_{in})$ and $h_{in}^{is}(P_{in})$ are respectively the isentropic specific volume and internal enthalpy provided using the second condition $s_{in} = s_0$. Both are a function of pressure and depend on the equation of state. The isentropic relationships are written using the SG EOS:
\begin{equation}
    \label{eqn:isentropicvolspec}
    v_{in}^{is}(P_{in}) = v_0 \left(\frac{P_0 + P_{\infty}}{P_{in} + P_{\infty}}\right)^{\frac{1}{\gamma}}
\end{equation}

\begin{equation}
    \label{eqn:isentropicenthalpy}
    h_{in}^{is}(P_{in}) = \frac{\gamma v_0 (P_0 + P_{\infty})}{\gamma-1} \left(\frac{P_0 + P_{\infty}}{P_{in} + P_{\infty}}\right)^{\frac{1-\gamma}{\gamma}} + e_{ref}
\end{equation}

In these equations, the quantities related to the inlet cross section, such as $P_{in}$, are unknown.

\subsection{Reference solutions : determining flow behavior}
\label{sec:ref_solutions}

The studies done on heated ducts in \cite{shapiro1953dynamics} and \cite{fathalli2018} completed by the previous observations have shown that once the duct is connected to a tank, as its inlet must be subsonic due to the coupling, it can only reach the sonic point at most and cannot become supersonic.
However, when appending a cooled subsection to the duct, we have seen that under a specific pressure ratio, the sonic state is in fact reachable and the cooled subsection can lead to  a supersonic state at  the outlet flow.

In this part, we find out by analytical calculations what are the conditions required to determine the flow regime.

As introduced in the previous observations, this type of description may remind adiabatic nozzle flows, with the following analogy done: 
\begin{compactitem}
    \item The converging part of the nozzle would be the heated subsection of the rectilinear duct.
    \item The diverging part would be the cooled subsection. 
    \item The fixed geometry of the nozzle (ie: the area ratio between the throat and the outlet $A^*/A(x)$) is viewed similarly to the fixed power function $Q(x)$ applied to the duct.
    \item The throat of the nozzle would be the point of heat flux change ($x_{heat}$). 
\end{compactitem}

We will then apply this analogy and we first hypothesize the existence of the same theoretical critical pressure ratios presented for example in \cite{shapiro1953dynamics}: $\Pi_{1}, \Pi_{2}$ and $\Pi_{3}$ (table~\ref{tab:Rpc}). They divide the flow into different specific regimes: fully subsonic, presence of a sonic point (choked flow), supersonic, shocked... Identifying those critical pressure ratios will allow us to categorize and properly determine the flow behavior. It must be noted that despite not yet having noticed a steady shock wave in the cooled subsection ($\Pi_{1} > \Pi \geq \Pi_{2}$), its existence is alleged and may be proven true or untrue later on.

\begin{table}
    \begin{center}
    \def~{\hphantom{0}}
    \begin{tabular}{c l}
    \toprule
    $\Pi = 1$ & No flow \\
    
    $1 > \Pi > \Pi_{1}$ & Fully subsonic flow \\
    
    $\Pi = \Pi_{1}$ & Subsonic inlet, choked flow at the end of the heated subsection, subsonic outlet \\
    
    \multirow{2}{*}{$\Pi_{1} > \Pi > \Pi_{2}$} & Subsonic inlet, choked flow at the end of the heated subsection, steady shock wave\\& in cooled subsection, subsonic outlet \\
    
    \multirow{2}{*}{$\Pi = \Pi_{2}$} & Subsonic inlet, choked flow at the end of the heated subsection, steady shock wave\\& at the outlet cross section \\
    
    \multirow{2}{*}{$\Pi_{2} > \Pi > \Pi_{3} $} & Subsonic inlet, choked flow at the end of the heated subsection, supersonic outlet\\& (Shock waves outside of the duct) \\ 
    
    \multirow{2}{*}{$\Pi = \Pi_{3}$} & Subsonic inlet, choked flow at the end of the heated subsection, supersonic outlet \\&(Adapted flow outside of the duct) \\
    
    \multirow{2}{*}{$\Pi_{3} > \Pi$} & Subsonic inlet, choked flow at the end of the heated subsection, supersonic outlet \\&(Expansion waves outside of the duct) \\
    \bottomrule
    \end{tabular}
    \caption{Nozzle pressure ratios definition applied to heated and cooled duct.}
    \label{tab:Rpc}
    \end{center}
\end{table}

The different values of critical $\Pi$ are now sought by solving the system of equations composed of the \textit{local solution} and the \textit{tank-inlet} relations.

\subsubsection{Fully subsonic flow}

The flow is supposed to be fully subsonic when $\Pi > \Pi_1$ ($M<1$ throughout the whole duct). Given that we do not reach sonic state at the end of the heated subsection, the duct also remains subsonic in the cooled subsection. Therefore, information coming from further than the point of heat flux variation (such as the outlet state) can impact the inlet.

{\bf Inlet state :} Firstly, we seek for the determination of the inlet state, as it will constraint the rest of the flow. In the subsonic case, the outlet pressure $P_{out}$ is a known constant and pressure information coming from downstream has an impact on the inlet flow of the duct. Equation~(\ref{eqn:rayleigh}) is evaluated between $(\cdot)_{in}$ and $(\cdot)_{out}$ to numerically determine the inlet pressure $P_{in}$ under the following implicit form:

\begin{equation}
\label{eqn:fPin_sub}
    f(P_{in}) = \dot m_s^2(v_{out} - v_{in}^{is}) + P_{out} - P_{in} = 0
\end{equation}

On one hand, the isentropic relations between the tank and the inlet allow to calculate the specific volume at the inlet $v_{in}^{is}(P_{in})$ with equation~(\ref{eqn:isentropicvolspec}) and the mass flow rate $\dot m_s(P_{in})$ with equation~(\ref{eqn:massflowrateinlet}). On the other hand, the local relations evaluated between the inlet and the outlet allow the determination of the specific volume at the outlet $v_{out}(P_{in})$, with equation~(\ref{eqn:v_crussard}), as the heat flux $Q_{out} \varpropto \dot q_{s,out}$ is also known.

The value of the inlet Mach number $M_{in}(P_{in})$ is determined by the isentropic relation between the tank and the inlet, through equation~(\ref{eqn:stagnantpressure}).

{\bf Duct state :} Once the inlet state is known, we now seek to characterize the flow at any point $x$ in the duct. This will be done by using the local solutions of the duct. The ratio of stagnation temperature at point $x$, denoted $T_{0,x}/T_{0,in}$, is obtained through equation~(\ref{eqn:T0x/T0in}). This equation is coupled with equation~(\ref{eqn:T0_ratio}), and both express the so-called $T_0$-change. This yields the following equation, which is a biquadratic function of $M_x$:

\begin{equation}
    \label{eqn:polyMach}
    f(M_x) = M_x^4 \left(\beta B - \gamma^2 A \frac{T_{0,x}}{T_{0,in}} \right) 
    + M_x^2 \left(B - 2\gamma \frac{T_{0,x}}{T_{0,in}} \right)
    - A \frac{T_{0,x}}{T_{0,in}} = 0, 
    \begin{cases}
        \beta = \frac{\gamma-1}{2} \\
        A(P_{in}) = M_{in}^2 \left(1+\beta M_{in}^2 \right) \\
        B(P_{in}) = (1+\gamma M_{in}^2)^2
    \end{cases}
\end{equation}

See appendix \textit{Solving $M_x$} for solutions of $f(M_x)=0$: there are 4 solutions, 2 of which are negative and thus non-physical, and 2 other which are positive, thus physical, but either subsonic ($<1$) or supersonic ($>1$). As we are studying the fully subsonic case, the lowest positive root of the equation ($0<M_x^{sub}<1$) represents the subsonic Mach number of this case's flow.

Given the Mach number at any point $x \in [0;L]$, the pressure $P_x$ is computed from equation~(\ref{eqn:PMach}), with $M_{in}$ determined from equation~(\ref{eqn:stagnantpressure}): 

\begin{equation}
    \label{eqn:Pi}
    P_{x}(P_{in}) = (P_{in} + P_{\infty}) \frac{1 + \gamma M_{in}^2}{1 + \gamma M_x^2} - P_{\infty}
\end{equation}

\subsubsection{Choked flow}

The flow is choked once $\Pi \leq \Pi_1$: a sonic state is reached in the duct. This state can only be observed at the $x_{heat}$ location, which is the end of the heated subsection. Given that the flow is sonic at $x_{heat}$, no information can come from downstream of this location: it acts a barrier for any backwards travelling wave. Then, the behavior of the flow within the cooled subsection can follow different paths, depending on the critical pressure ratios presented in table~\ref{tab:Rpc}. Our goal now is to determine these hypothetical values.

{\bf Inlet state :} Given  prior considerations, to determine the inlet state in the case of a choked flow, we focus on the the coupled system \textit{tank-inlet} and \textit{inlet-}$x_{heat}$ instead of \textit{inlet-outlet}. Indeed, the flow is sonic at $x_{heat}$ and that will not change.

Equation (\ref{eqn:rayleigh}) is therefore evaluated between $(\cdot)_{in}^*$ and $x_{heat} \equiv (\cdot)_{heat}$ to numerically determine $P_{in}^*$, where $(\cdot)_{in}^*$ denotes the inlet state for which the flow ends up choked:

\begin{equation}
    \label{eqn:fPin_sonic}
    f(P_{in}^*) = \dot m_s^2(v_{heat} - v_{in}^{is}) + P_{heat} - P_{in}^* = 0
\end{equation}

In the same fashion as the subsonic flow, the relations between the tank and the inlet allow to compute the mass flow rate $\dot m_s(P_{in}^*)$ with equation~(\ref{eqn:massflowrateinlet}) and the specific volume $v_{in}^{is}(P_{in}^*)$ with equation~(\ref{eqn:isentropicvolspec}). The local relations evaluated between $(\cdot)_{in}^*$ and $(\cdot)_{heat}$ lead to the determination of the specific volume $v_{heat}(P_{in}^*)$, through equation~(\ref{eqn:v_crussard}), given that $Q_{heat} \varpropto \dot q_{s,heat}$ is known and fixed.

The remaining unknown of equation~(\ref{eqn:fPin_sonic}) is the pressure at the end of the heated subsection, $P_{heat}$. Our hypothesis that the flow is choked means that it reaches the sonic state and thus, $M_{heat} = 1$. The expression connecting the Mach number and the pressure is given by equation~(\ref{eqn:PMach}), evaluated locally between $(\cdot)_{in}^*$ and $(\cdot)_{heat}$:

\begin{equation}
    \label{eqn:Pheat}
    P_{heat}(P_{in}^*) = (P_{in}^* + P_{\infty}) \frac{1 + \gamma M_{in}^2}{1 + \gamma} - P_{\infty}
\end{equation}

The value of the inlet Mach number $M_{in}(P_{in}^*)$ is determined by the isentropic relation between the tank and the inlet, via equation~(\ref{eqn:stagnantpressure}).

{\bf Duct state :} Now, the inlet state is known and from it we seek to characterize the flow at any point $x$ of the duct. The determination of the stagnation temperature $T_{0,x}$ through equation~(\ref{eqn:stagnantenth}), coupled with the ratio of stagnation temperature in the duct $T_{0,x}/T_{0,in}$ with equation~(\ref{eqn:T0_ratio}), yields the same polynomial of order 4 presented by equation~(\ref{eqn:polyMach}), function of Mach number at point $x$ denoted $M_x(P_{in}^*)$. See appendix for the resolution of the polynomial.

In the heated subsection ($x \in [0;x_{heat}[$), the Mach number is necessarily subsonic. Thus, its value the lowest positive root of equation~(\ref{eqn:polyMach}) ($0<M_x^{sub}(P_{in}^*) \leq 1$). We also know that the sonic point is reached at ($x =x_{heat}$).
In the cooled subsection however ($x \in ]x_{heat};L]$), once the flow is choked, there are two possible regimes depending on the pressure ratio that is imposed: subsonic (therefore still $0<M_x^{sub}(P_{in}^*) < 1$), or supersonic (therefore $M_x^{sup}(P_{in}^*) > 1$). They respectively correspond to the lowest positive root and highest positive root of equation~(\ref{eqn:polyMach}).

This behavior can be seen in figure~\ref{fig:T0ratio_big}: once the beginning of the cooled subsection is sonic, there are two possible states for $T_{0,x}/T_{0,in} < 1$: one where $M_x > 1$ and one where $M_x < 1$.

Depending on Mach number value $M_x(P_{in}^*)$, the pressure for each possible paths is evaluated from equation (\ref{eqn:Pi}). Again, the value of Mach number at inlet $M_{in}(P_{in}^*)$ is determined from equation~(\Ref{eqn:stagnantpressure}).

\begin{figure*}
\centering
    \begin{subfigure}{0.49\textwidth}
        \includegraphics[width=\textwidth]{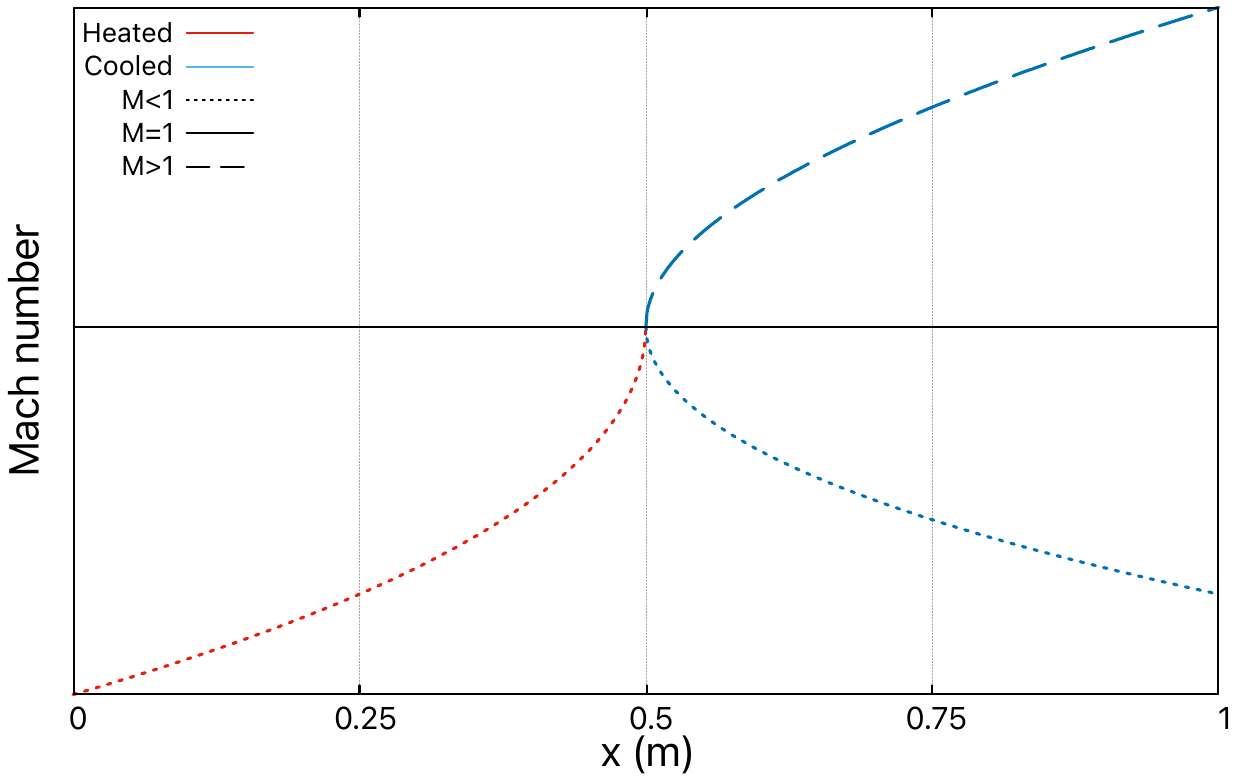}
        \label{fig:rpc_mach}
    \end{subfigure}
    \begin{subfigure}{0.49\textwidth}
        \includegraphics[width=\textwidth]{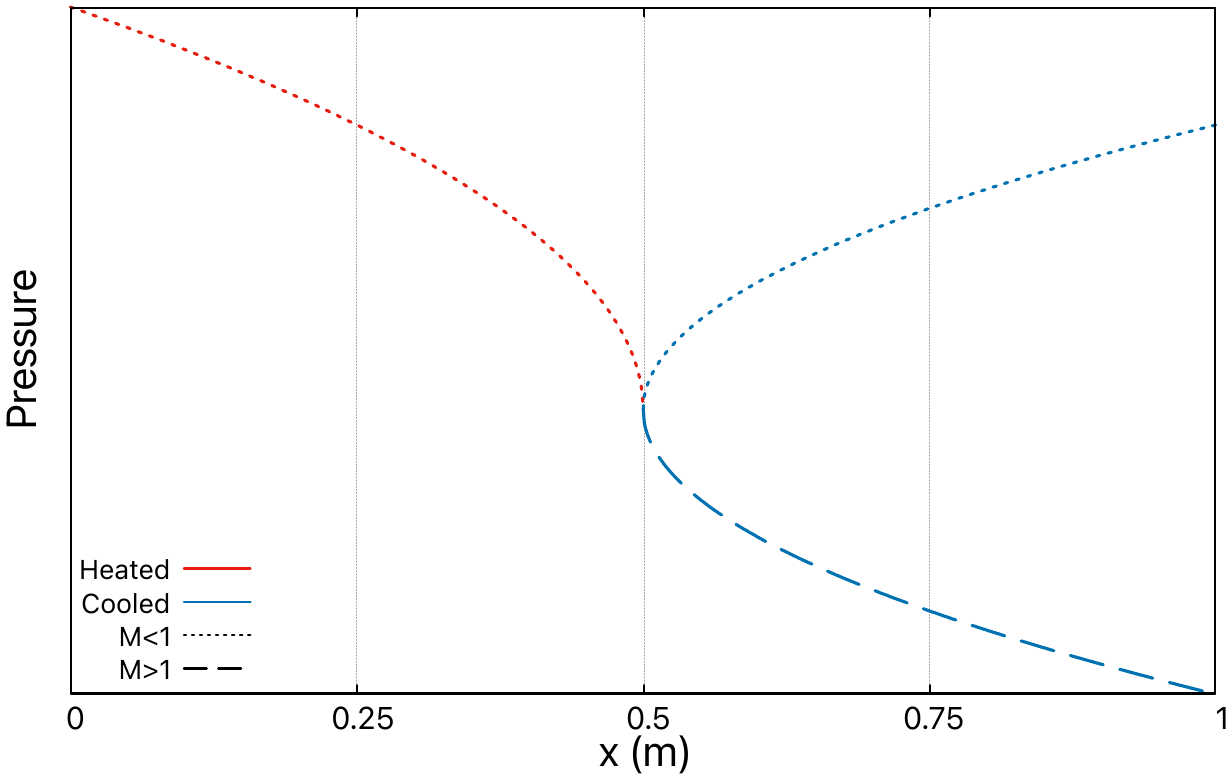}
        \label{fig:rpc_pressure}
    \end{subfigure}
    \caption{Mach number (left) and pressure (right) evolution for a choked flow within a duct. The heated subsection (red line) starts subsonically and reaches the sonic point ($M_x \leq 1$, dotted line), but two possible paths exist within the cooled subsection (blue lines), depending on $\Pi$: subsonic ($M_x < 1$, dotted line) or supersonic ($M_x > 1$, dashed line). }
    \label{fig:rpc_paths}
\end{figure*}

Figure~\ref{fig:rpc_paths} depicts the possible critical paths described above.

{\bf Critical outlet states :} When the flow is choked (sonic point at the end of the heated subsection), one can see that the outlet pressure $P_{out}$ is not needed anymore. Indeed, $P_{out}$ is absent from the system formed by equations \eqref{eqn:fPin_sonic}, \eqref{eqn:Pheat} and \eqref{eqn:polyMach} to determine $M_x$. In this configuration, only the heat flux function and the tank parameters condition the flow through the variation of stagnation temperature described with equation~(\ref{eqn:T0x/T0in}).

The critical pressures for choked flow and then subsonic outlet and supersonic outlet are evaluated from equation~(\ref{eqn:Pi}) at point $(\cdot)_{out}$. Relating back to table~\ref{tab:Rpc}, they respectively allow to determine $\Pi_{1} = P_{out}^{sub}/P_0$ and $\Pi_{3} = P_{out}^{sup}/P_0$, both functions of $P_{in}^*$.

To determine $\Pi_{2}=P_{out}^{shock}/P_0$, we assume an adiabatic shock wave located in the outlet section of the duct, starting from the supersonic outlet state. Being adiabatic, the shock wave state remains on the \textit{Hugoniot curve} of the supersonic outlet state and must end up at the intersection with the \textit{Rayleigh line}, as in a choked flow, its slope $-\dot m_s^2$ is fixed by heated duct properties.

\begin{figure*}
    \centering
    \includegraphics[width=0.85\textwidth]{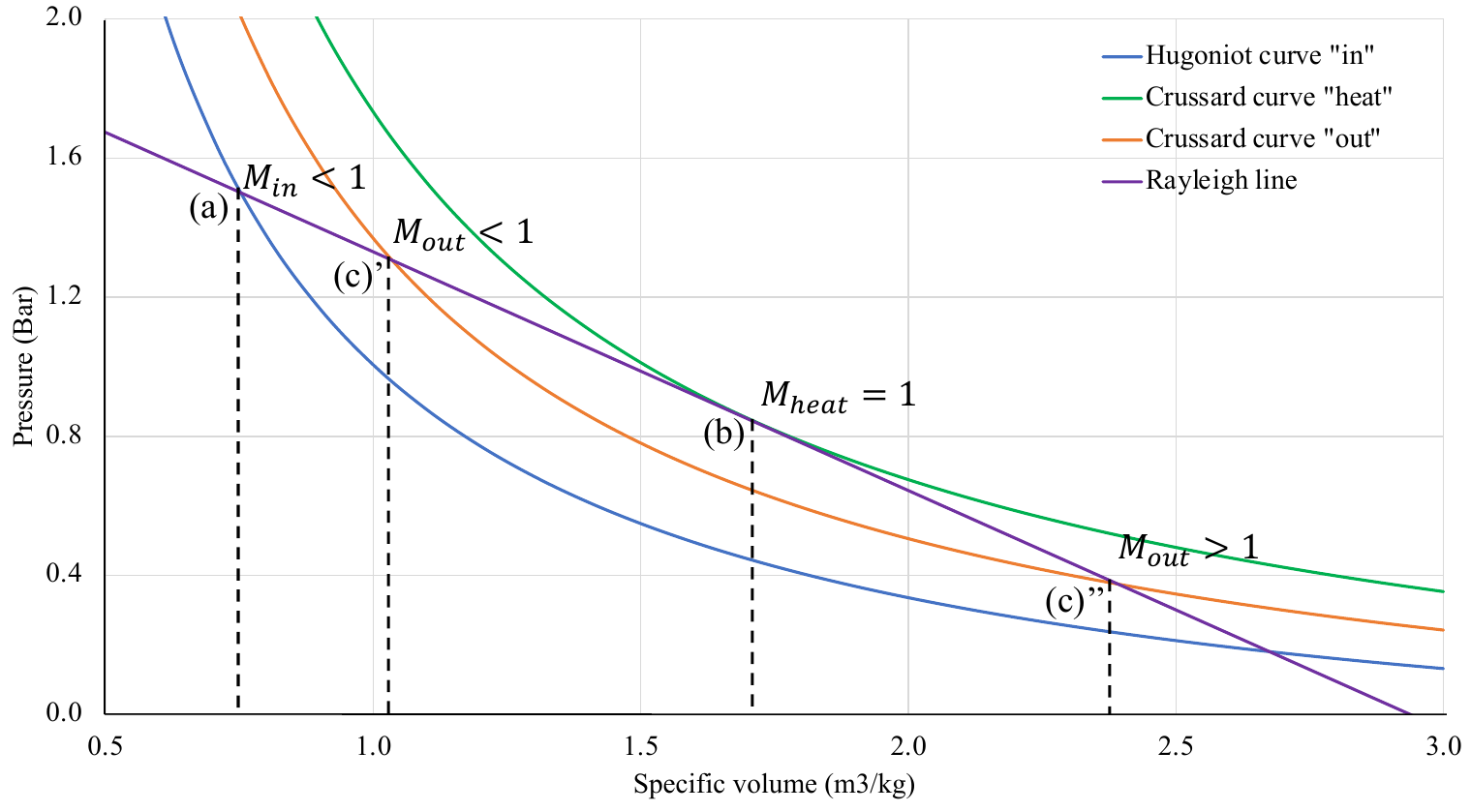}
    \caption{Thermodynamic path for a heated/cooled choked flow in the $(P,v)$ plane, with corresponding Mach number values.
    Starting at a subsonic state on the Hugoniot curve $"in"$ \textbf{(a)}, the fluid receives a certain amount of heat.
    If the pressure ratio is critical in the given configuration, the flow accelerates until it reaches sonic point at $"heat"$ state. The corresponding Crussard curve $"heat"$ is tangent to the Rayleigh line of the flow at the sonic point \textbf{(b)}. 
    Then, as the fluid is cooled down to reach outlet state, it moves to another Crussard curve $"out"$ \textbf{(c)}: the outlet state winds up at the intersection of the Rayleigh line and the outlet Crussard curve.
    Depending on the $\Pi$, the outlet state is either the subsonic state \textbf{(c')}, or the supersonic state \textbf{(c'')} ; there can be no intermediate steady shocked state between the latters. }
    \label{fig:PV_Rpc2}
\end{figure*}

Constrained by these two curves, the outlet shock wave can only bring the supersonic outlet state to the subsonic outlet state. This means that the shocked outlet state is equivalent to the subsonic outlet state determined previously with $\Pi_{1}$ ($P_{out}^{shock} = P_{out}^{sub} \Rightarrow \Pi_{2} = \Pi_{1}$). The thermodynamic path is displayed in figure~\ref{fig:PV_Rpc2}.

Thus recalling table~\ref{tab:Rpc}, given that $\Pi_{2} = \Pi_{1}$, we have demonstrated that there can be \textbf{no steady shock wave in the cooled subsection}. If any part of this subsection has reached a supersonic state, in a steady state the whole subsection will become supersonic. In the choked flow conditions, given a specific $Q_{heat}$ and inlet state, only two outlet states are possible.

The updated pressure ratio table is displayed in table~\ref{tab:Rpc_updated}.

\begin{table}
    \begin{center}
    \def~{\hphantom{0}}
    \begin{tabular}{c l}
        \toprule
        $\Pi = 1$ & No flow \\
        
        $1 > \Pi > \Pi_{1}$ & Fully subsonic flow \\
        
        $\Pi = \Pi_{1}$ & Subsonic inlet, choked flow at the end of the heated subsection, subsonic outlet \\
        
        \multirow{2}{*}{$\Pi_{1} > \Pi > \Pi_{3}$} & Subsonic inlet, choked flow at the end of the heated subsection, supersonic outlet\\ &(Shock waves outside of the duct) \\
         
        \multirow{2}{*}{$\Pi = \Pi_{3}$} & Subsonic inlet, choked flow at the end of the heated subsection, supersonic outlet \\&(Adapted flow outside of the duct) \\
        
        \multirow{2}{*}{$\Pi_{3} > \Pi$} & Subsonic inlet, choked flow at the end of the heated subsection, supersonic outlet \\&(Expansion waves outside of the duct) \\ 
        \bottomrule
    \end{tabular}
    \caption{Pressure ratios definition for heated and cooled duct.}
    \label{tab:Rpc_updated}
    \end{center}
\end{table}

At this point, knowing the heating and cooling powers $Q_{heat}$ and $Q_{cool}$, the critical pressure ratios are known. We can then determine the flow behavior within the duct for a given pressure ratio $\Pi$.

Circling back to the nozzle studies analogy, only knowing the ratio of exit and throat areas is necessary (therefore only one data), whereas here a couple of data is needed (any combination of $Q_{heat}$ and $Q_{cool}$).

\section{Critical pressure ratios behavior}
\label{sec:rpc_analysis}

In the following section, we study the critical pressure ratios: the goal is to determine the limits of the possible regimes, depending on the power function $Q(x)$ that the duct is subjected to, and provide analytical and physical limitations of our study framework. We especially take a look at the behavior of $\Pi_1$, as it marks the threshold between a subsonic outlet and a supersonic outlet.

\subsection{Asymptotic analysis - Limit cases}

\subsubsection{Isentropic flow} To begin with, when the flow in the duct is neither heated nor cooled (thus isentropic), we deduce that the state within the duct is constant. As $P(x) = P_{out} \left. \right. \forall x$, the Mach number through the whole duct is determined by equation~(\ref{eqn:stagnantpressure}). If the flow actually reaches sonic state, then $P_{in}^* = P_{out}$ and $M_{in} = M_{out} = 1$. Equation (\ref{eqn:stagnantpressure}) evaluated in this specific configuration yields the unheated isentropic sonic limit case:

\begin{equation}
    \label{eqn:Pi_is}
    \Pi(Q(x)=0 , \forall x) = \dfrac{P_{out}}{P_0} = \dfrac { P_0 + P_{\infty} }{P_0} \left( \dfrac{2}{\gamma+1} \right) ^{\frac{\gamma}{\gamma-1}} - \dfrac { P_{\infty} }{P_0}  \equiv \Pi_{is}
\end{equation}

In the case of ideal gases, $\Pi_{is} = \left( \dfrac{2}{\gamma +1} \right)^{\frac{\gamma}{\gamma-1}}$.

\subsubsection{Heated flow} If the duct is now heated, the flow state is no longer constant. If it reaches sonic point at its outlet, it means that $M_{out}=1$. The outlet pressure $P_{out}$ can be expressed by evaluating equation (\ref{eqn:PMach}) between $(\cdot)_{in}$ and $(\cdot)_{out}$:

$$
    P_{out} = (P_{in}^* + P_{\infty}) \frac{1 + \gamma M_{in}^2}{1 + \gamma} - P_{\infty}
$$

If the heating power is largely increased, the inlet pressure $P_{in}^*$ rises to reach tank pressure $P_0$ and the pressure along the duct decreases as the flow is heated, to reach the outlet state determined by the pressure ratio. This can be seen through figure~\ref{fig:observations}. A quasi zero pressure gradient between the tank and the inlet implies that the flow is almost at rest at the inlet. The inlet pressure cannot get higher than the tank pressure, as the direction of the flow would then be reversed and our study hypotheses would no longer be valid. Through isentropic equation~(\ref{eqn:stagnantpressure}) between the tank and the inlet, we deduce the inlet Mach number $M_{in} \rightarrow 0$. In that configuration, we can express the theoretical value of lowest critical pressure ratio $\Pi_h$: 

\begin{equation}
    \label{eqn:Pih}
     \Pi(Q \rightarrow +\infty) = \dfrac{P_{out}}{P_0} = \dfrac{P_0-\gamma P_\infty} {(1+\gamma)P_0} \equiv \Pi_h 
\end{equation}

\begin{figure*}
\centering
    \includegraphics[width=0.85\textwidth]{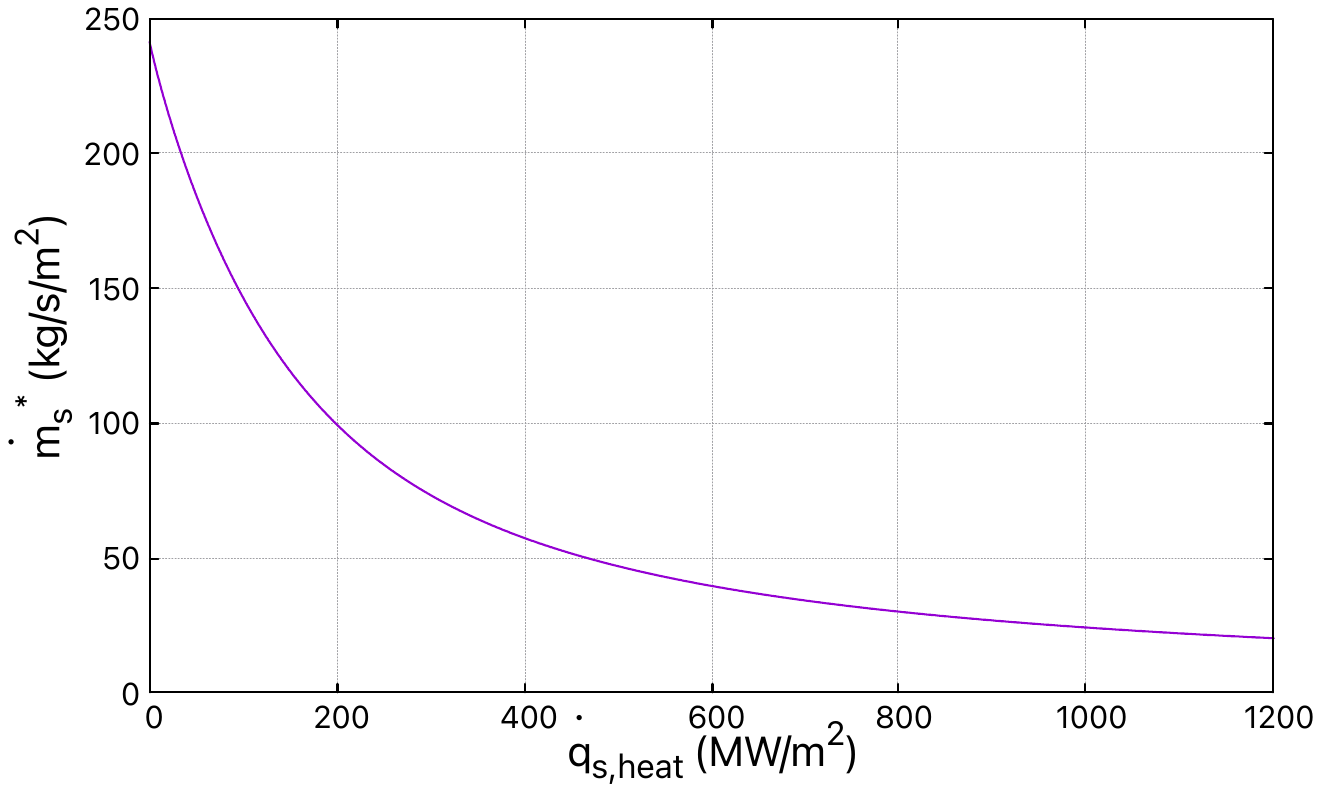}
    \caption{$\dot m_s^*$ evolution depending on heating power. When $\dot q_{s,heat}$ increases, $\dot m_s^*$ decreases.}
    \label{fig:ms_crit_f_qsheat}
\end{figure*}

In the case of ideal gases, $\Pi_h = 1/(1+\gamma)$. However, for SG, this limit depends on the stagnation pressure $P_0$ (note for both cases, the limit case  is $\Pi_h < 1$).
For a given heating power, once the duct reaches sonic state, it has been shown that the mass flow rate is fixed. Increasing the heating power leads to a decrease of critical mass flow rate ($\dot m_s^* \equiv \dot m_s(P_{in}^*)$), as shown by figure~\ref{fig:ms_crit_f_qsheat}.

\begin{figure*}
\centering
    \includegraphics[width=0.85\textwidth]{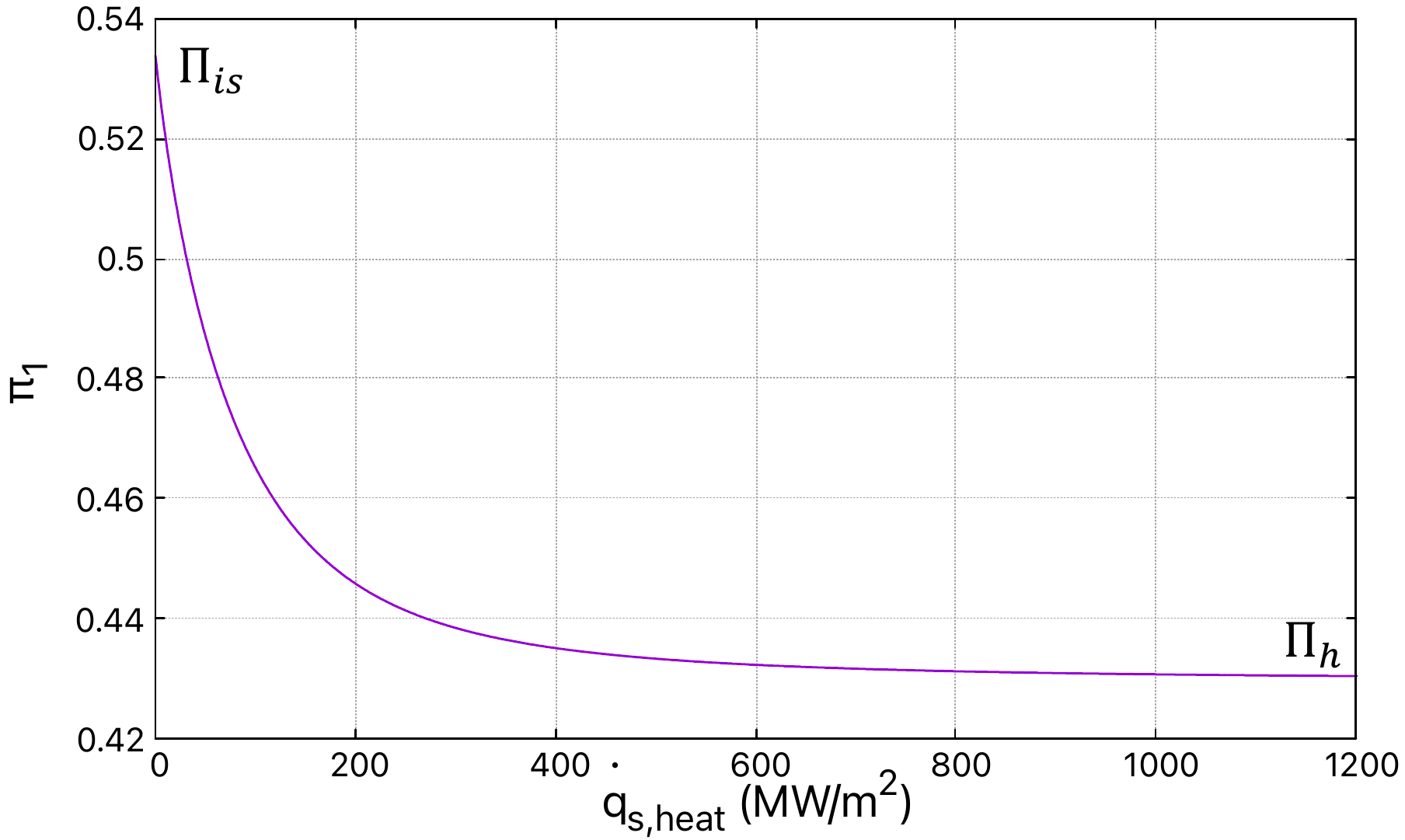}
    \caption{$\Pi_1$ evolution depending on heating power. When $\dot q_{s,heat} \rightarrow 0 W/m^2$, $\Pi_1 \rightarrow \Pi_{is}$, and when $\dot q_{s,heat} \rightarrow +\infty$, $\Pi_1 \rightarrow \Pi_h$.}
    \label{fig:rpc_fqsheat_qscool0}
\end{figure*}

Figure~\ref{fig:rpc_fqsheat_qscool0} displays the behavior of $\Pi_1$ versus  the heat flux $\dot q_{s,heat} \propto Q_{heat}$ in order to remove geometrical dependency, in the following conditions:

\begin{compactitem}
    \item Gas: $\gamma=1.358,\, C_v=1247,\, e_{ref}=1.97\cdot 10^7 \text{J/kg},\, P_\infty=0$ Pa
    \item Tank state: $P_0 = 1.5$ Bar, $T_0 = 394$ K
\end{compactitem} 

When the heating power decreases, $\Pi_1$ converges towards the singular value $\Pi_{is}$ presented in equation~(\ref{eqn:Pi_is}). It is the pressure ratio needed to obtain a choked flow without any heating in a given configuration : if $\Pi > \Pi_{is}$, the flow can never reach the sonic point, as heating will only decrease the Mach number. On the other hand, if the heating power increases, the value of $\Pi_1$ decreases and tends towards the theoretical pressure ratio $\Pi_h$ introduced with equation~(\ref{eqn:Pih}). This corresponds to the lowest pressure that can be reached at the sonic point when increasing the heating power. At a certain point, increasing heating becomes pointless, as the critical pressure ratio $\Pi_1$ is bounded from below by the value of $\Pi_h$.

\subsubsection{Heated and cooled flow} Heated and cooled flows are here considered. We assume that the flow is choked and therefore reaches sonic point at the end of the heated subsection, thus $M_{heat}=1$ and $P_{heat}$ is expressed by equation~(\ref{eqn:Pheat}).

On one hand, if the cooling power is negligible in front of the heating power, the flow behavior does not vary much after entering the cooled subsection, as one could consider the cooled subsection close to adiabatic. The pressure ratio needed to bring the end of the heated subsection to the sonic point follows the curve presented by figure~\ref{fig:rpc_fqsheat_qscool0}. If the heating power is high enough, the pressure ratio tends towards the value of the critical pressure ratio of a duct that is solely heated: in that case, $P_{out} \rightarrow P_{heat}$ and $\Pi_1 \rightarrow \Pi_h$.

On the other hand, when the cooling power applied to the flow is much larger than the heating power, and thus the flow is overall cooled down at the outlet, the outlet state pressure $P_{out}$ caused by a sonic flow increases compared to the inlet pressure (see figure~\ref{fig:PV}). However, to keep our hypothesis of flow going from a tank to an outlet, the pressure ratio $\Pi = P_{out}/P_0$ must be less than 1 ; otherwise, the direction of the flow will be reversed. Therefore, we pose the following inequality, which will yield a condition on the lowest applicable cooling power $Q_{cool}^{min}$: $ \Pi_1 = P_{out}/P_0 < 1 $.

To solve it, we must determine the outlet pressure $P_{out}(P_{in}^*)$ by evaluating equation~(\ref{eqn:Pi}) at the outlet. It depends on the inlet state, which is known and fixed by the properties of the heated subsection, as we suppose a choked flow. The inlet pressure $P_{in}^*$ is determined from equation~(\ref{eqn:fPin_sonic}) and the inlet Mach number $M_{in}(P_{in}^*)$ is determined from equation~(\ref{eqn:stagnantpressure}).

The outlet pressure also depends on the outlet Mach number $M_{out}^{sub}(P_{in}^*)$, which can be evaluated by finding the lowest positive root of equation~(\ref{eqn:polyMach}) at point $(\cdot)_{out}$. $M_{out}^{sub}$ depends on the outlet stagnation temperature $T_{0,out}$, which depends on the power applied $Q_{out} = Q_{heat} + Q_{cool}$. Therefore, after retrieving $T_{0,out}^{min}$, the lowest possible cooling power that keeps a flow going from the tank to the outlet can be easily determined from equation~(\ref{eqn:T0x/T0in}):

\begin{equation}
    \label{eqn:Qcool_min}
    Q_{cool}^{min} > \left( \frac{T_{0,out}^{min}}{T_{0,in}} - 1  \right) \dot m_s C_p T_{0,in} S - Q_{heat}
\end{equation}

Beyond the apparent simplicity of this equation, the calculation of $T_{0,out}^{min}$ is not trivial and equation details can be found in appendix \textit{Solving $Q_{cool}^{min}$}.

As we have seen with the above limit cases, when $Q_{heat} \gg |Q_{cool}|$, the outlet pressure tends towards $P_{heat} < P_{in}^*$. When $Q_{heat} \ll |Q_{cool}|$, the outlet pressure rises towards $P_0$ (at most, given our study hypothesis). Thus, when the system under study is both heated and then cooled down, $\Pi_1 \in ]\Pi_h;1[$. 

\subsection{Critical pressure ratios dependence regarding $Q_{heat}$ and $Q_{cool}$ (abacus)}

\begin{figure*}
\centering
    \includegraphics[width=0.85\textwidth]{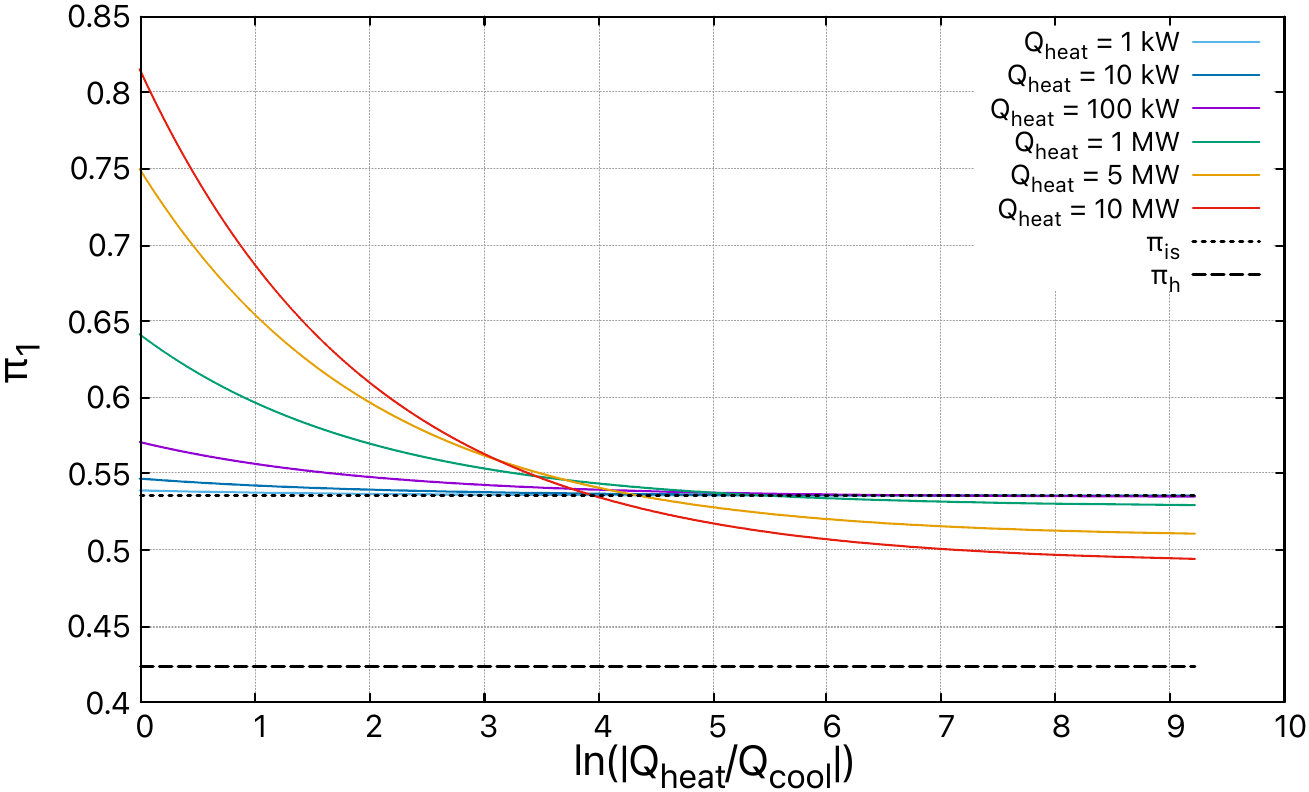}
    \caption{Critical pressure ratio $\Pi_1$ evolution depending on the value of ratio of heating over cooling power, for different values of heating power. For a given heating power and power ratio, if $\Pi > \Pi_1$, then the flow is fully subsonic. If $\Pi < \Pi_1$, then it is choked and supersonic in the cooled subsection.}
    \label{fig:fQheatQcool}
\end{figure*}

Now, we consider a nondescript combination of heating and cooling power. Figure~\ref{fig:fQheatQcool} shows the evolution of the critical pressure ratio $\Pi_1$ against the ratio of heating over cooling power $|Q_{heat}/Q_{cool}|$. For a fixed power ratio, there are multiple possible critical pressure ratios. Therefore, either the heating power or cooling power must be a fixed parameter as well. For example, for a power ratio $|Q_{heat}/Q_{cool}|=1$, if $Q_{heat}= 5\,MW$, then the outlet flow becomes supersonic for $\Pi < \Pi_1=0.75$.

Let us consider a fixed power ratio $|Q_{heat}/Q_{cool}|$, where either $Q_{heat} \gg |Q_{cool}|$ or $Q_{heat} \ll |Q_{cool}|$ \textcolor{red}.

On one hand, if the heating power $Q_{heat}$ decreases, the flow critical pressure ratios converge towards the singular value $\Pi_{is}$ presented in equation~(\ref{eqn:Pi_is}). It is the pressure ratio needed to obtain a choked flow without any heating in a given configuration : if $\Pi > \Pi_{is}$, the flow can never reach the sonic point, as heating will only decrease the Mach number.

On the other hand, if the heating power $Q_{heat}$ increases, the flow critical pressure ratios convergence point decreases and tends towards the theoretical pressure ratio $\Pi_h$ introduced with equation~(\ref{eqn:Pih}). This is the lowest pressure that can be reached at the sonic point when increasing the heating power. At a certain point, increasing heating becomes pointless, as the critical pressure ratio $\Pi_1$ is bounded from below by the value of $\Pi_h$.
    \section{Results}
\label{sec:results}

We present some results that illustrate the flow behaviour in various conditions: these results are obtained by the analytical solutions proposed in this study. They are also the answer to the question presented in the introduction: \textit{What are the possible solutions for a flow in a duct of length $L$ receiving a certain amount of power $Q_{out}$ ?} Then, we also present comparisons with numerical results provided by the numerical solution of the Euler equations system (\ref{eqn:euler}), thanks to a code based on finite volume method.

\subsection{Parameters study}

For every test case that is presented, we consider a duct of length $L=1\,$m  heated for $x \in [0;x_{heat}] $ and cooled for $ x \in ]x_{heat};L]$, with a total volume of $V=1/(4\pi)$ $\text{m}^3$. For a cylindrical duct, that leads to a radius of $R=1/(2\pi)\,$m, a cross-section $S=\pi R^2 = 1/(4 \pi)\,\text{m}^2$ and a perimeter $P=2 \pi R = 1\,$m. 
 
The tank conditions are also held constant in the different cases : $P_0 = 1.5$ Bar, $T_0 = 394$ K. 

The fluid under study is a gas obeying the IG EOS with the parameters $\gamma=1.358, C_v=1247, e_{ref}=1.97\cdot 10^7\,\text{J/kg}, P_{\infty}=0 $ (which is particular case of the SG EOS). The heat capacity at constant volume $C_v$ is required to extract the temperature given by relationship $e=C_v T$. 

The value of the heating power is $Q_{heat} = 200$ kW. The cooling power would be a parametric term to get some various conditions through $Q_{out} = Q_{heat}+Q_{cool}$. The corresponding value of $Q_{cool}^{min}$ is $-5.45$ MW.

All those parameters define the system under consideration. To calculate various flows, a dynamic parameter must be changed to differentiate these cases, which is what we will study thereafter.

\subsubsection{Variation of pressure ratio $\Pi$}

The dynamic parameter of this study is obviously the outlet pressure through the ratio $\Pi$: we take a look at the impact of the pressure ratio on the behavior of the flow.

\begin{figure*}
\centering
    \begin{subfigure}{0.49\textwidth}
        \includegraphics[width=\textwidth]{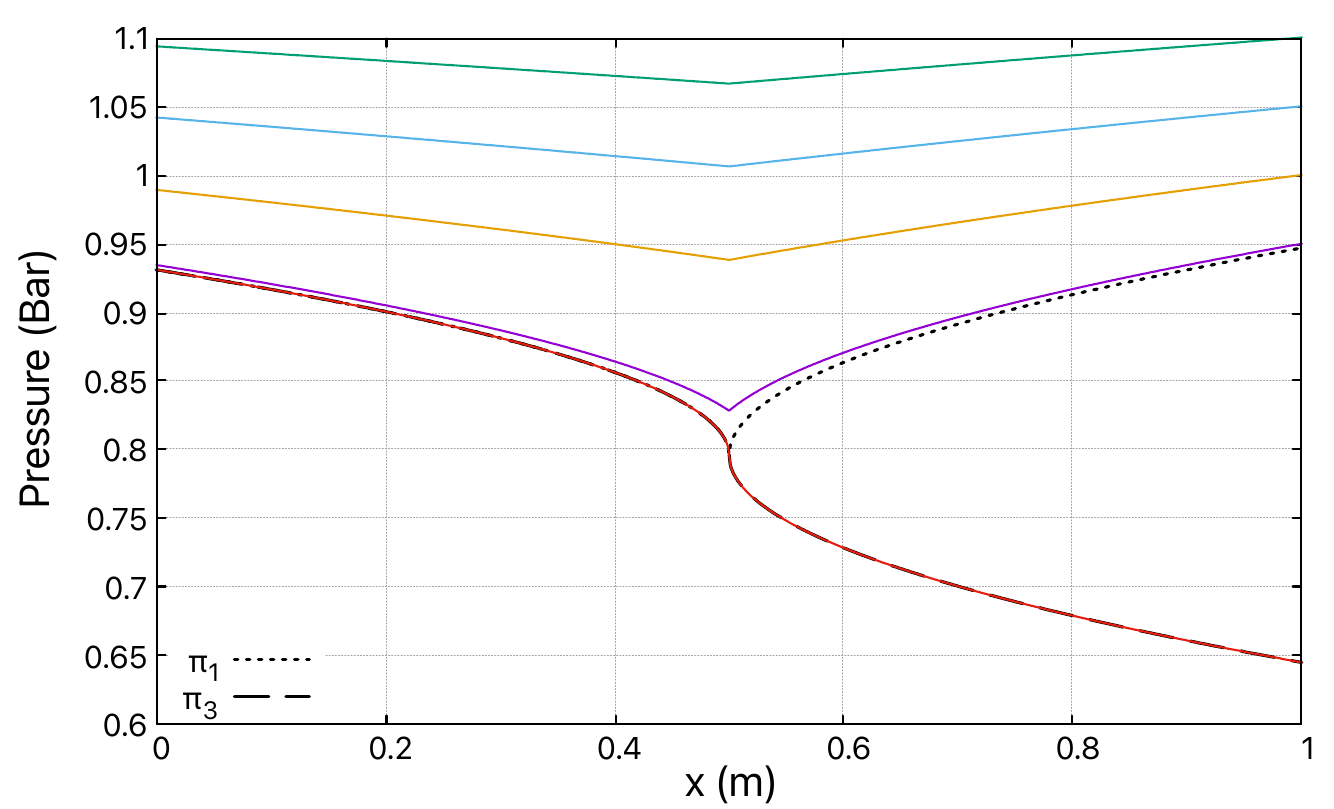}
        \label{fig:P_Rp_Qm50kW}
    \end{subfigure}
    \begin{subfigure}{0.49\textwidth}
        \includegraphics[width=\textwidth]{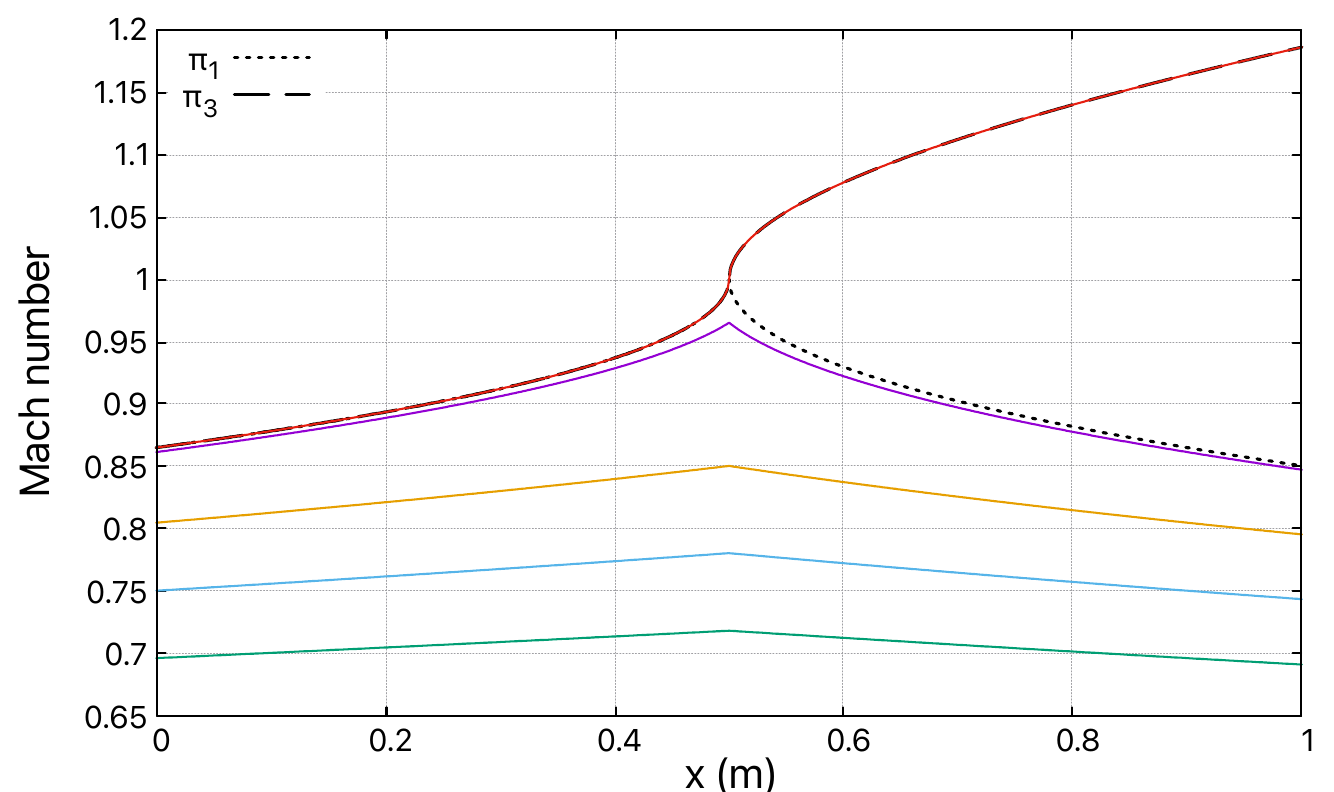}
        \label{fig:Mach_Rp_Qm50kW}
    \end{subfigure}
    \begin{subfigure}{0.49\textwidth}
        \includegraphics[width=\textwidth]{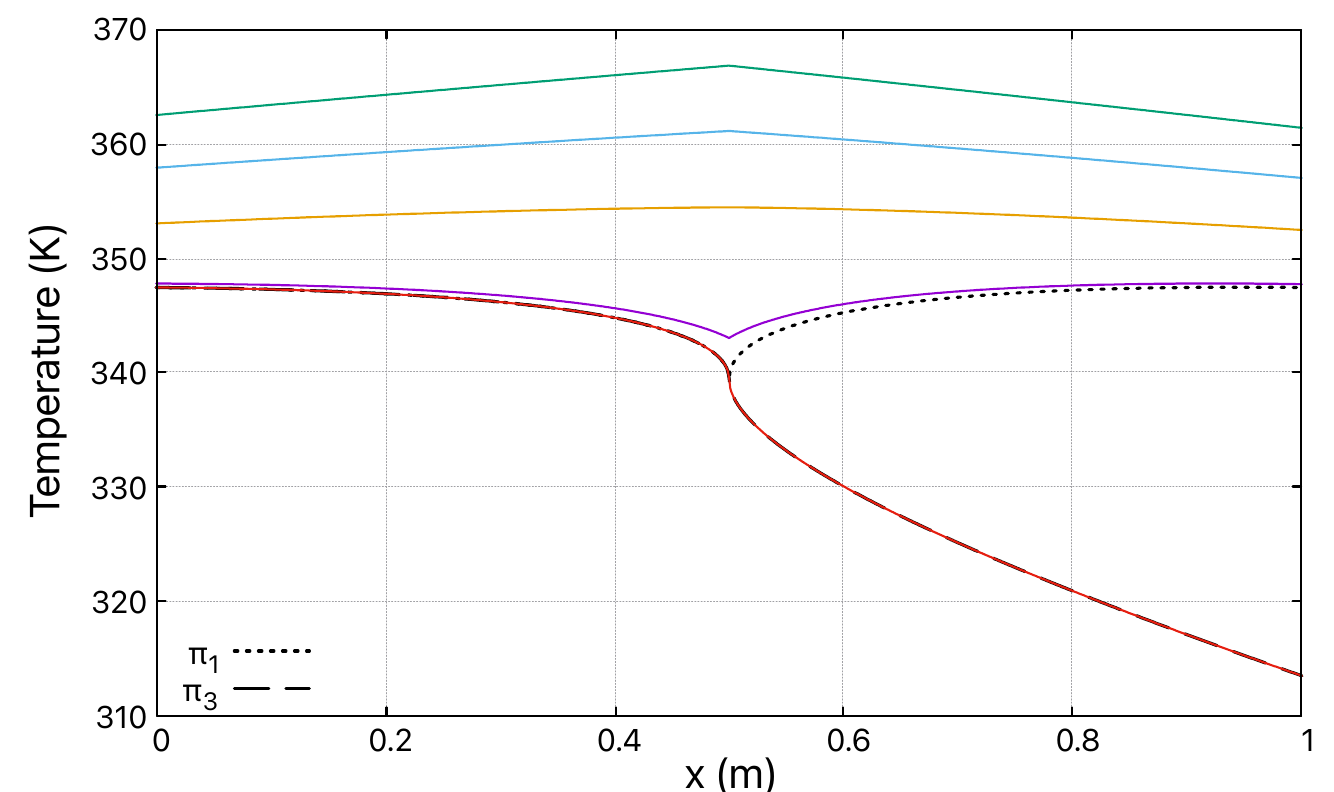}
        \label{fig:T_Rp_Qm50kW}
    \end{subfigure}
    \begin{subfigure}{0.49\textwidth}
        \includegraphics[width=\textwidth]{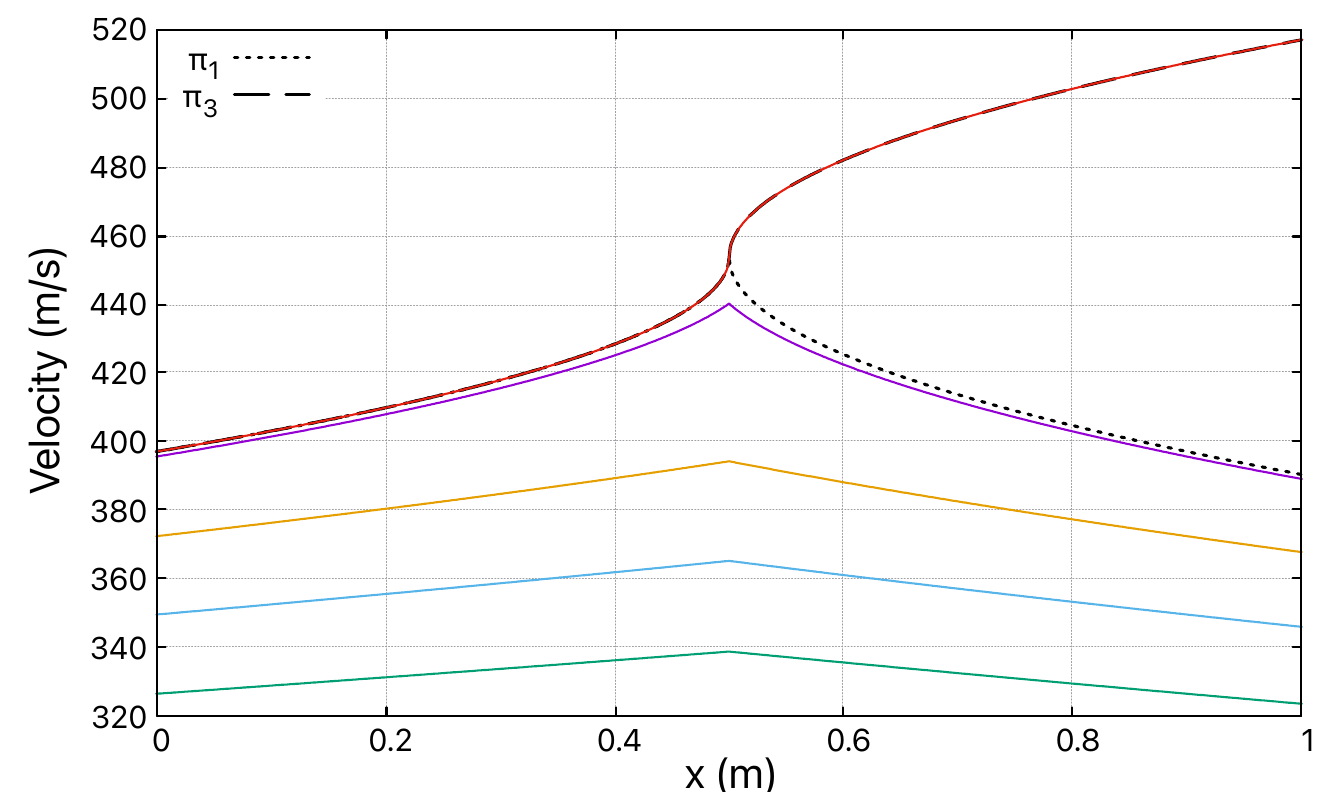}
        \label{fig:u_Rp_Qm50kW}
    \end{subfigure}
    \caption{Behavior of the flow variables depending on different pressure ratios, for a fixed couple $\{ Q_{heat}=200$ kW, $Q_{cool}=-250$ kW$\}$. As $\Pi$ decreases, the flow gets progressively closer to choking. Once $\Pi < \Pi_{1}$, the steady flow remains supersonic in the cooled subsection (red solution).}
    \label{fig:varRp}
\end{figure*}

In this test case, we pose $Q_{cool}= -250$ kW, meaning that $Q_{out}=-50$ kW: the flow is overall more cooled than heated. The critical pressure ratios can first be calculated by the method presented in Section \ref{sec:ref_solutions}, which yields the following values: $\Pi_1 = 0.6313$, $\Pi_3 = 0.4303$. The whole solution is then calculated between the inlet and the outlet. The particular solutions corresponding to $\Pi_1$ and $\Pi_3$ are labelled in figure~\ref{fig:varRp}. In this figure, the evolution of the pressure, the Mach number, the temperature and the flow velocity is depicted, for various values of $\Pi \in [0.60;0.73]$.

\begin{figure*}
\centering
    \includegraphics[width=0.85\textwidth]{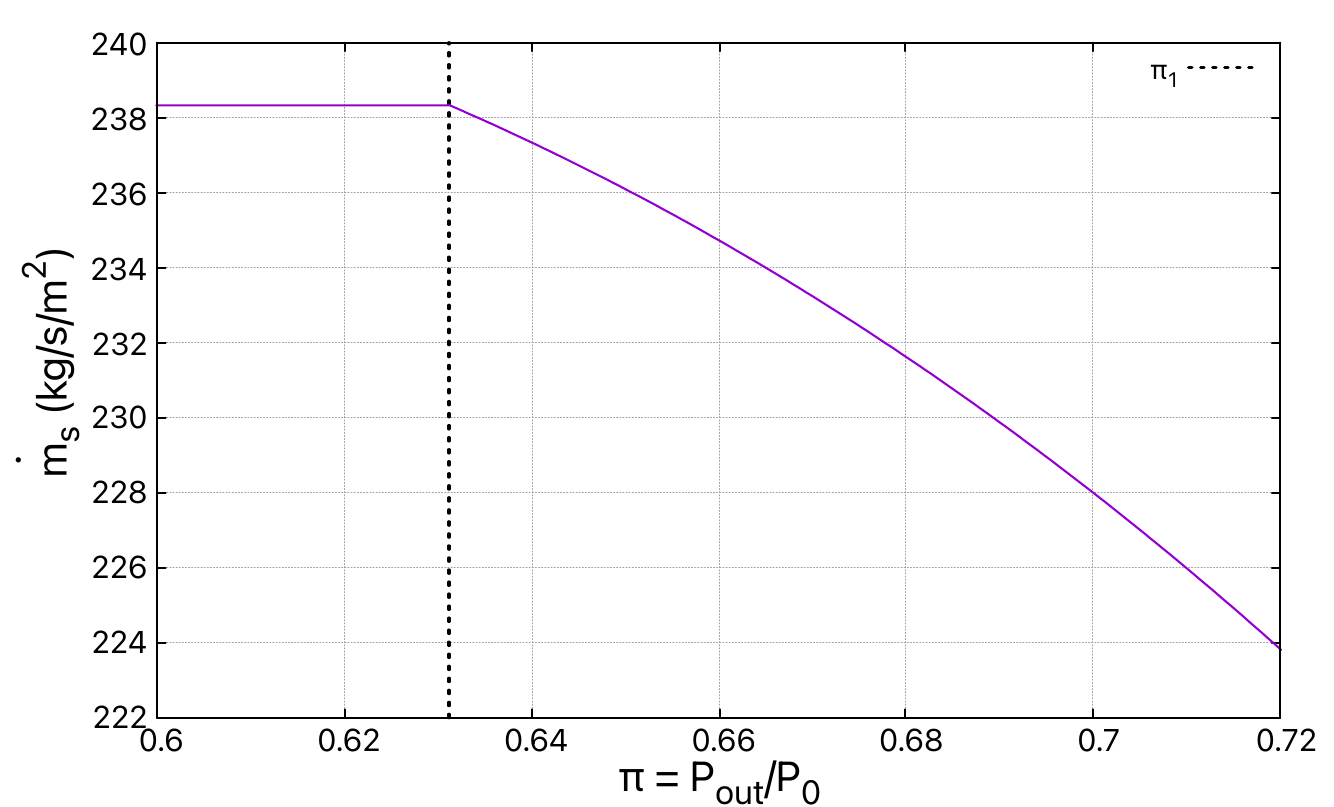}
    \caption{$\dot m_s$ evolution depending on pressure ratio $\Pi$ for a fixed couple $\{ Q_{heat}=200$ kW, $Q_{cool}=-250$ kW$\}$. As long as $\Pi > \Pi_1$, when the pressure ratio decreases, the flow accelerates and the mass flow rate increases. When $\Pi< \Pi_1$, $\dot m_s\equiv \dot m_s^*$ no longer varies as the flow is choked.}
    \label{fig:ms_fPi}
\end{figure*}

Several solutions are represented for $\Pi \ge \Pi_1$: the mass flow rate $\dot m_s(P_{in})$ varies as the state of the heated subsection still depends on the variations of $P_{out}$ (as depicted on figure~\ref{fig:ms_fPi}). In this range, the pressure and the Mach number follow a coherent pattern. The Mach number increases up to $x_{heat}$ and then decreases in the cooled part of the duct, and the pressure follows the reverse pattern. Extremum values are reached at $x_{heat}$ for $\Pi = \Pi_1$.

Once $\Pi < \Pi_1$ and the flow is choked, $\dot m_s(P_{in}^*)$ is fixed by heated subsection, which is determined and cannot change (see figure~\ref{fig:ms_fPi}). There is no steady shock wave in the cooled subsection and every supersonic solution is overlapping (see red curve in figure~\ref{fig:varRp}).

Let's take a look at the evolution of the temperature, which is quite interesting and relates an example of possible particular behaviors within compressible flows. For greater values of $\Pi$, in the heated part, the temperature increase is monotonous up until $x_{heat}$ and then, in the same manner, decreases in the cooled part. This result seems to be obvious: when a fluid is heated, its temperature increases. However, for lower values of $\Pi$, the temperature decreases in the heated part of the duct and increases in the cooled part (for $\Pi > \Pi_1$). This result is a known particularity of non-adiabatic compressible flows: the flow follows this pattern if $1/\sqrt\gamma < M < 1$ \cite{shapiro1953dynamics} (see for example purple curve in figure~\ref{fig:varRp}).

\subsubsection{Variation of outlet heat power $Q_{out}$}

We now fix the pressure ratio $\Pi$ and take a look at the impact of variable power.

As $Q_{heat}$ is fixed, modifying $Q_{out}$ is equivalent to modifying the cooling power $Q_{cool}$. When $Q_{out} > 0$, then $|Q_{cool}| < Q_{heat}$: therefore, there is overall more heating than cooling power on the duct. When $Q_{out} < 0$, then $|Q_{cool}| > Q_{heat}$: then, there is more cooling power.

\begin{figure*}
\centering
    \begin{subfigure}{0.49\textwidth}
        \includegraphics[width=\textwidth]{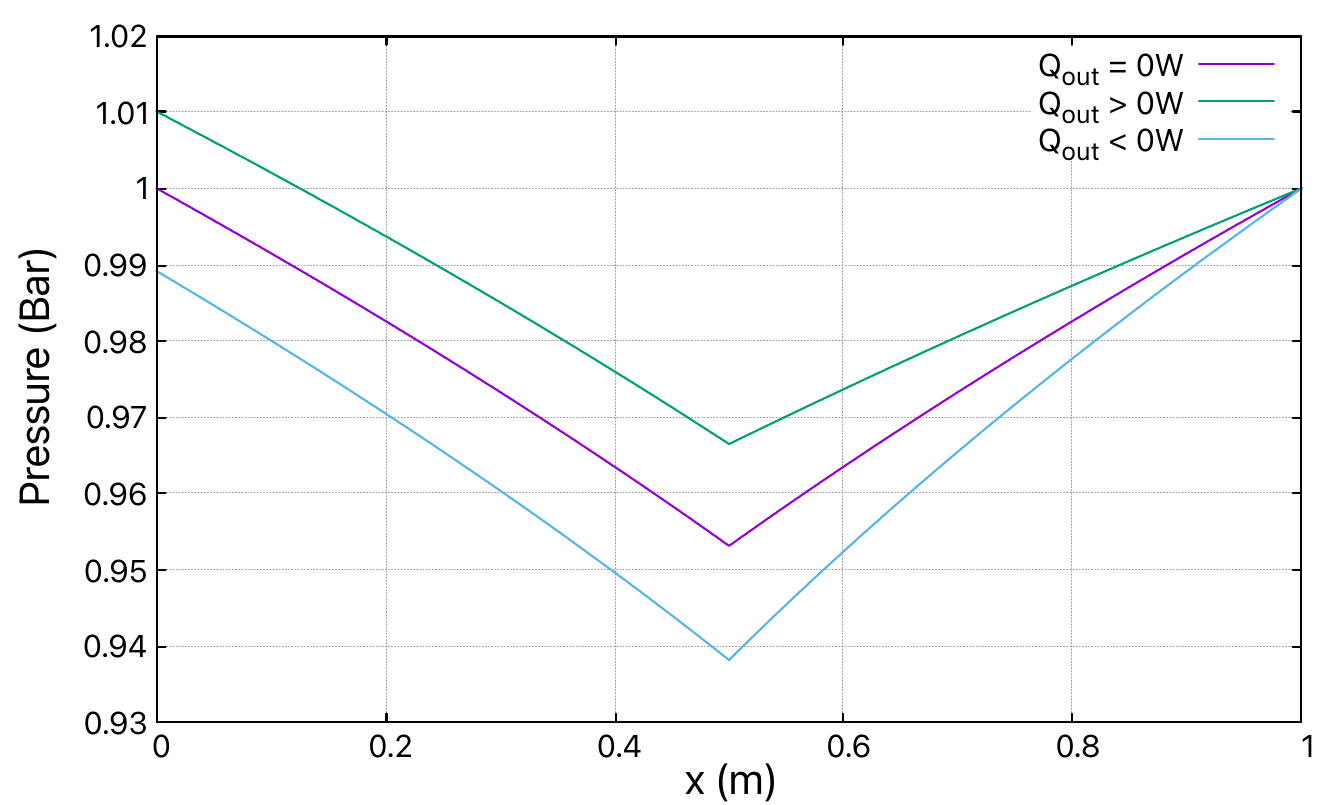}
        \label{fig:P_Rp_Rpc1}
    \end{subfigure}
    \begin{subfigure}{0.49\textwidth}
        \includegraphics[width=\textwidth]{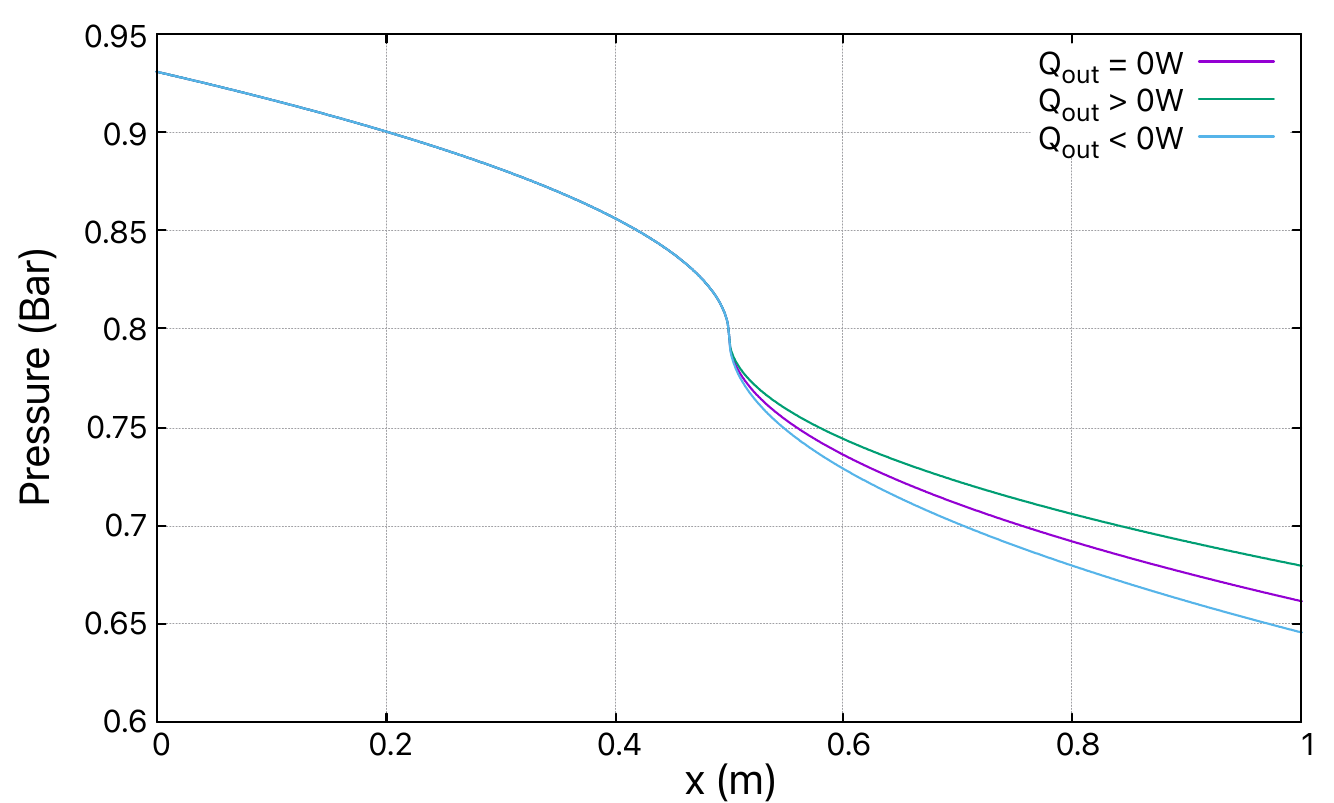}
        \label{fig:P_Rp_Rpc3}
    \end{subfigure}
    \begin{subfigure}{0.49\textwidth}
        \includegraphics[width=\textwidth]{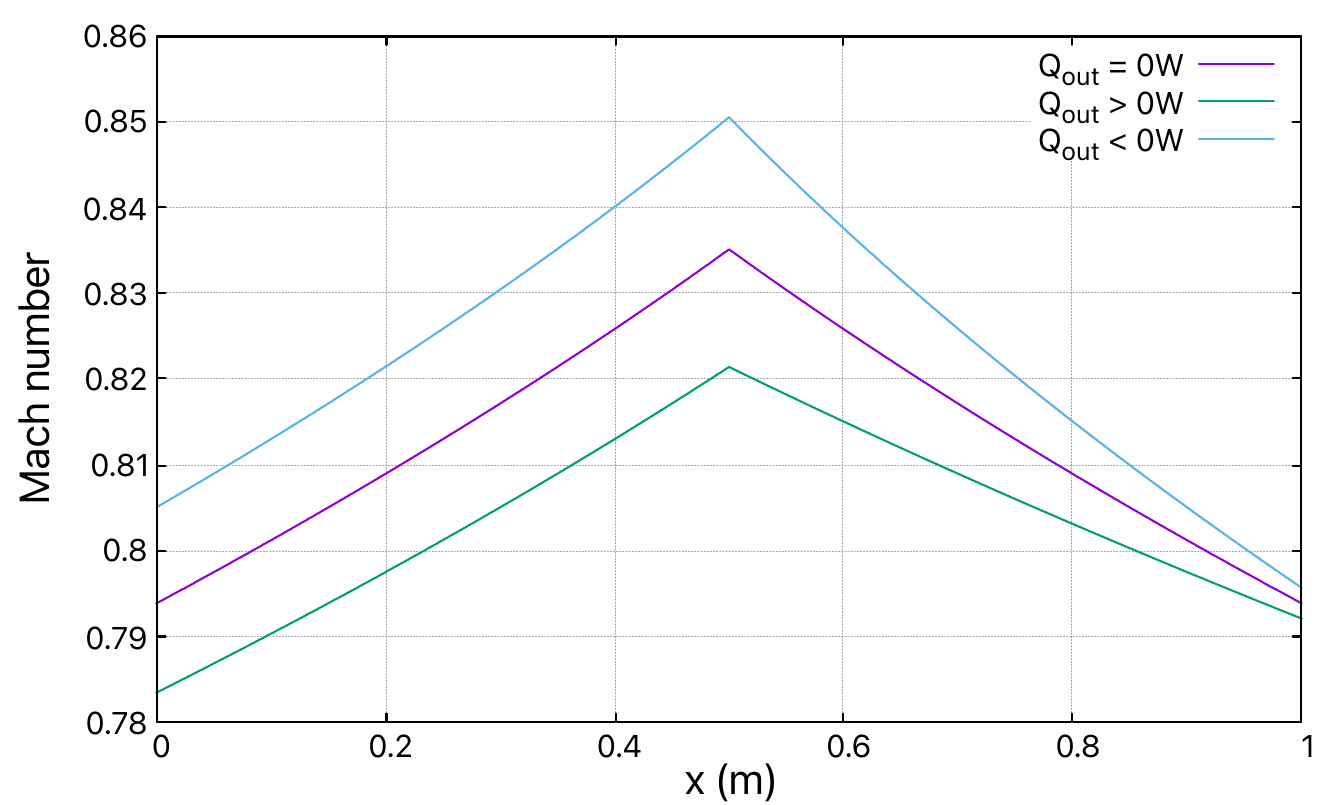}
        \label{fig:M_Rp_Rpc1}
    \end{subfigure}
    \begin{subfigure}{0.49\textwidth}
        \includegraphics[width=\textwidth]{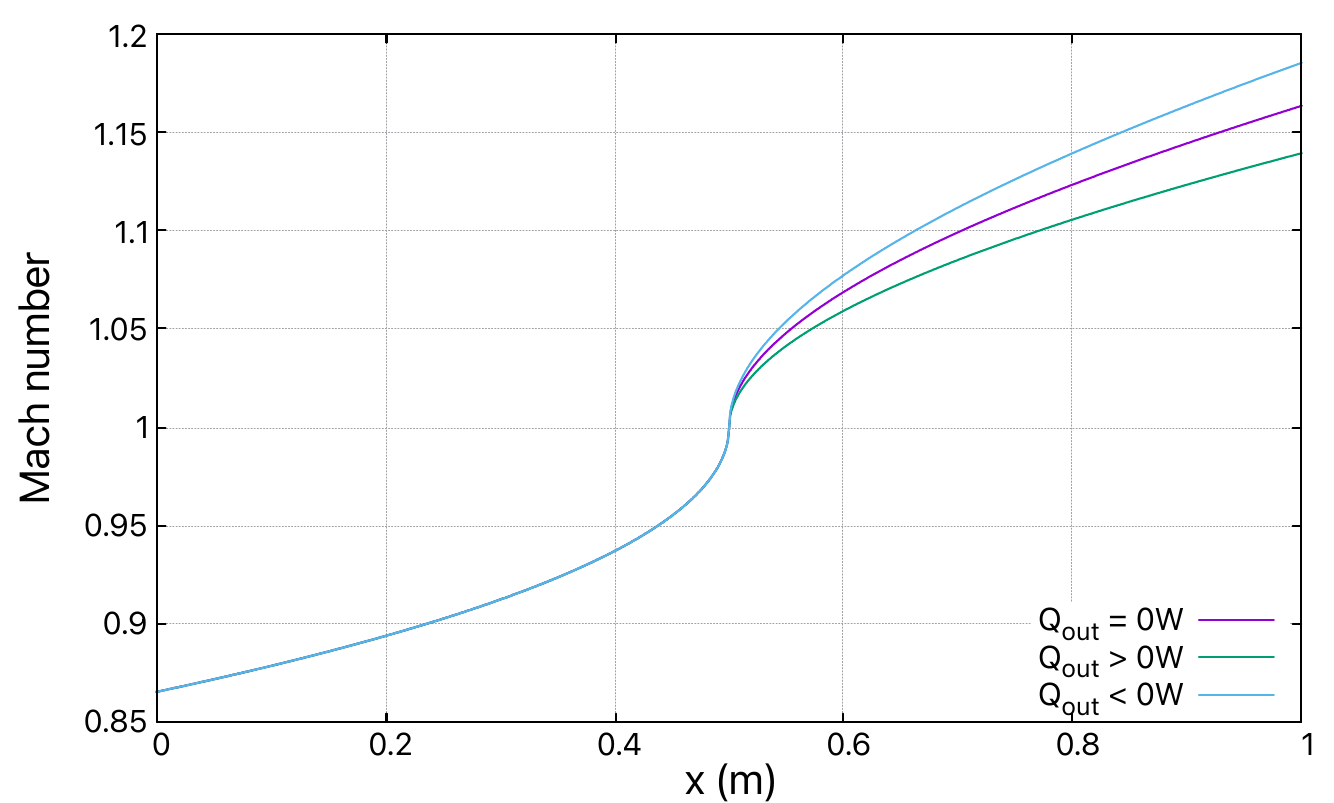}
        \label{fig:M_Rp_Rpc3}
    \end{subfigure}
    \caption{Pressure (top row) and Mach number (bottom row) evolution in a heated and cooled duct for different outlet states, with a fixed heating power. 
    For $\Pi > \Pi_1$ (left), the outlet state is constrained but the whole duct re-adapts.
    For $\Pi < \Pi_1$ (right), the heated subsection state is fixed but the cooled subsection re-adapts.}
    \label{fig:varRp_compa}
\end{figure*}

Figure~\ref{fig:varRp_compa} underlines the difference in flow behaviour for different values of outlet power $Q_{out}$, with a fixed $Q_{heat}$, depending on the flow regime. Overall, in the case of a non-adiabatic outlet, the pressure varies compared to an adiabatic outlet. Only two representative flow cases are considered: the first case is for a pressure ratio $\Pi > \Pi_1$, ensuring a fully subsonic flow, and the second case is for $\Pi < \Pi_1$.

For a subsonic flow ($\Pi > \Pi_1$), in the case of an adiabatic outlet, there is symmetry between inlet and outlet, which is lost for a non-adiabatic outlet. If the outlet state is no longer adiabatic, as the outlet pressure $P_{out}$ conditions the flow, the full duct re-adapts depending on the value of $Q_{out}$. Within the duct, as the flow is heated and cooled down but remains subsonic, it follows the behavior described by figure~\ref{fig:T0ratio_focus}. Between the inlet and the outlet, when $Q_{out} > 0$, the pressure generally decreases ($P_{in} > P_{out}$) and the Mach number generally increases ($M_{in} < M_{out}$). On the opposite, for $Q_{out}<0$, the pressure generally increases ($P_{in} < P_{out}$) and the Mach number generally decreases ($M_{in} > M_{out}$). 

For a supersonic outlet flow ($\Pi < \Pi_1$), the heated subsection is conditioned by $P_{heat}$, which itself is determined by the fixed heating power $Q_{heat}$. Thus, only modifying $Q_{out}$ cannot impact the heated subsection and the inlet state, but does change the cooled subsection state. When $Q_{out} > 0$, $P_{out}$ decreases less than when $Q_{out} < 0$, and the Mach number increases less. This can be explained by figure~\ref{fig:T0ratio_big}: now that the supersonic state is reached in the cooled subsection, heating leads to a decrease of Mach number and cooling leads to an increase of it.

There would be little information in studying the effects of the  value of $Q_{heat}$, which would only change the levels without changing the qualitative evolution of quantities, compared to the case presented in this section. We can simply underline the following points: depending on if the flow is choked or not, an increase of heating power will vary the inlet Mach number and the inlet pressure, as described in figure~\ref{fig:T0ratio_focus}. It will also decrease the value of $\Pi_1$, as figure~\ref{fig:fQheatQcool} shows.

\subsection{Example of validation for CFD codes}

These analytical results are used to validate CFD codes simulations of compressible single-phase flows in non-adiabatic ducts, whether they are being solely heated, cooled down, or both. In this subsection, we use the CFD code described in \cite{schmidmayer2018}, which uses the \cite{Godounov:1979ul} method, in the following configuration:

\begin{compactitem}
    \item Tank state: $P_0 = 1.5$ Bar, $T_0 = 394$ K
    \item Geometry: $V=1/4\pi\,\text{m}^3, x_{heat} = L/2$
    \item Power: $Q_{heat} = 200$ kW, $Q_{cool}=-250$ kW
\end{compactitem} 

We compute two different fluids: a gas ($P_\infty=0$) and a liquid ($P_\infty \neq 0$):

\begin{compactitem}
    \item Gas: $\gamma=1.358,\, C_v=1247,\, e_{ref}=1.97\cdot 10^7\, \text{J/kg},\, P_\infty=0\,\text{Pa}$
    \item Liquid: $\gamma=3.423,\, C_v=1231.2,\, e_{ref}=-1.15\cdot 10^6\, \text{J/kg},\, P_\infty=10^4\,\text{Pa}$
\end{compactitem}

\begin{figure*}
\centering
    \begin{subfigure}{0.49\textwidth}
        \includegraphics[width=\textwidth]{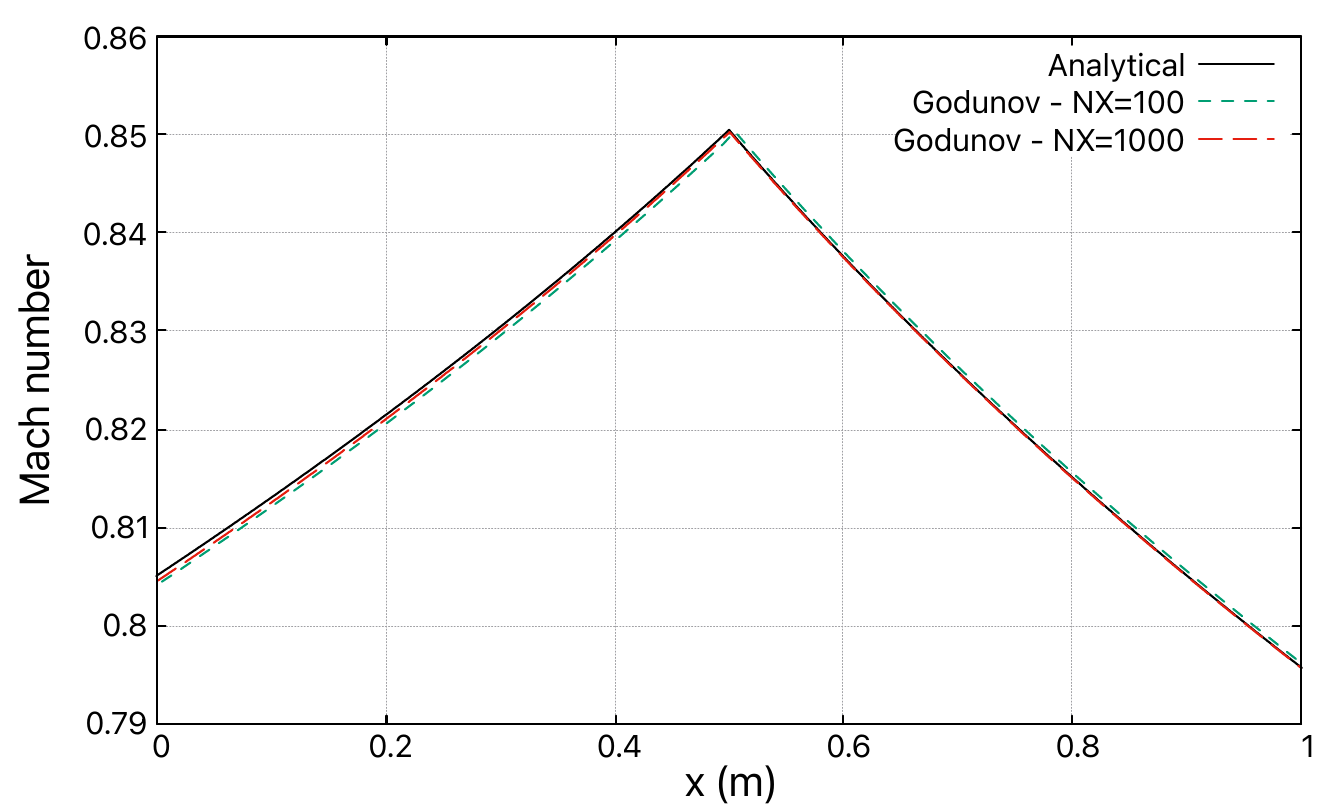}
        \label{fig:godu_varNX_sub_M_vapor}
    \end{subfigure}
    \begin{subfigure}{0.49\textwidth}
        \includegraphics[width=\textwidth]{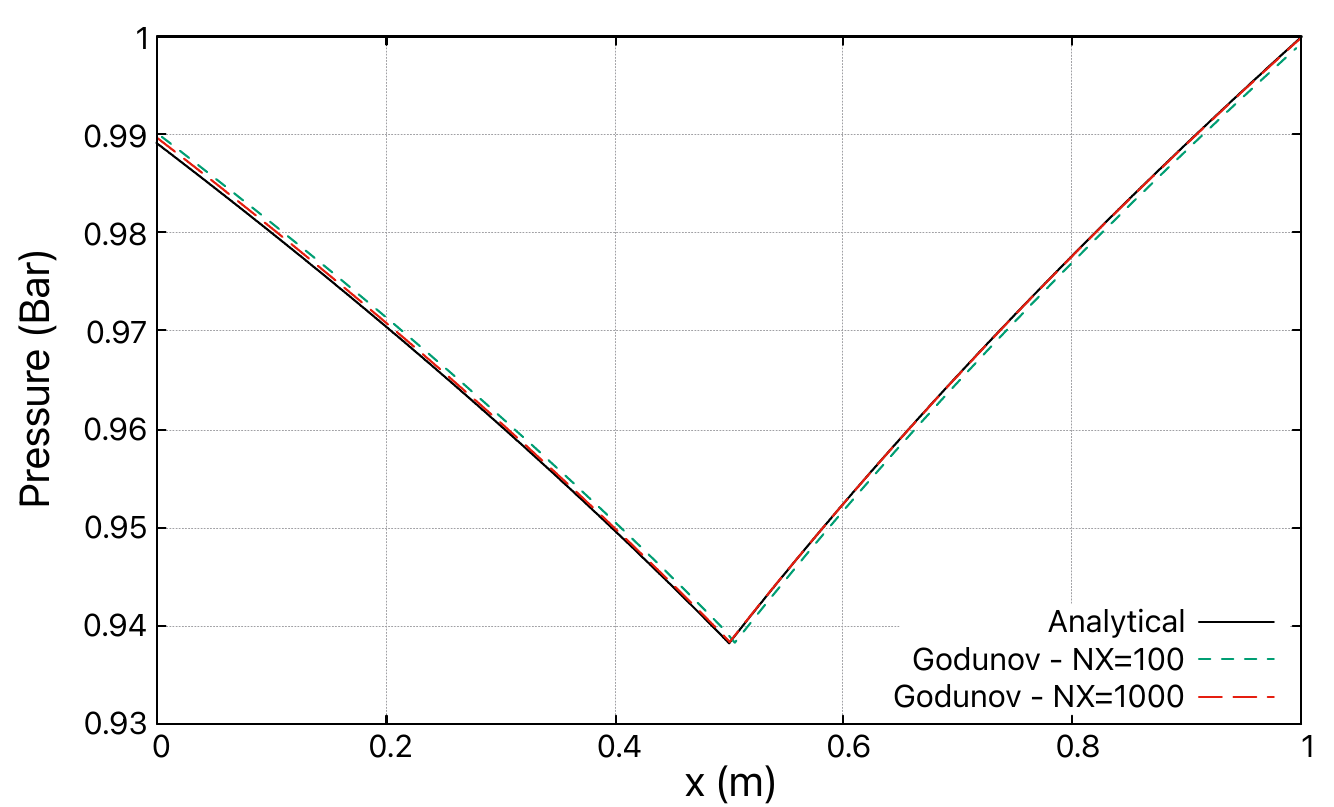}
        \label{fig:godu_varNX_sub_P_vapor}
    \end{subfigure}
    \begin{subfigure}{0.49\textwidth}
        \includegraphics[width=\textwidth]{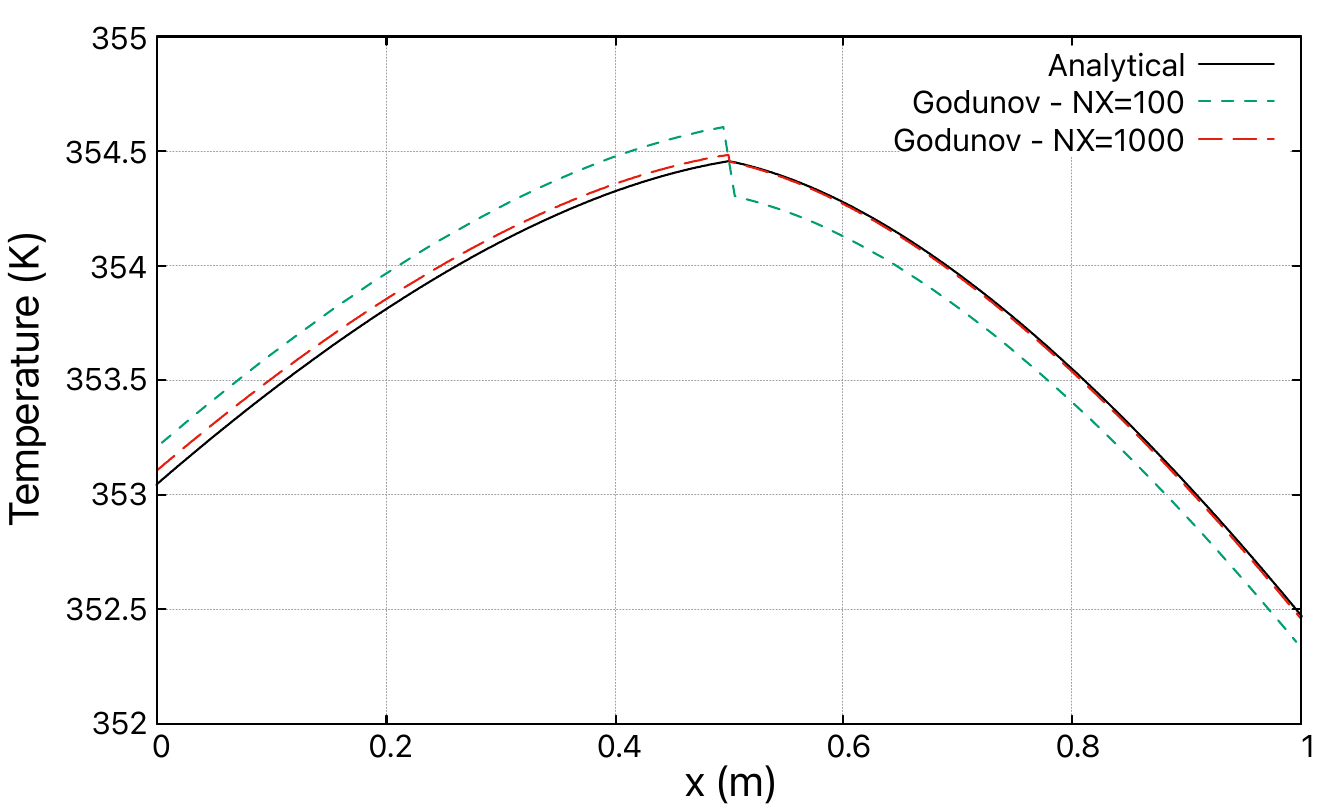}
        \label{fig:godu_varNX_sub_T_vapor}
    \end{subfigure}
    \begin{subfigure}{0.49\textwidth}
        \includegraphics[width=\textwidth]{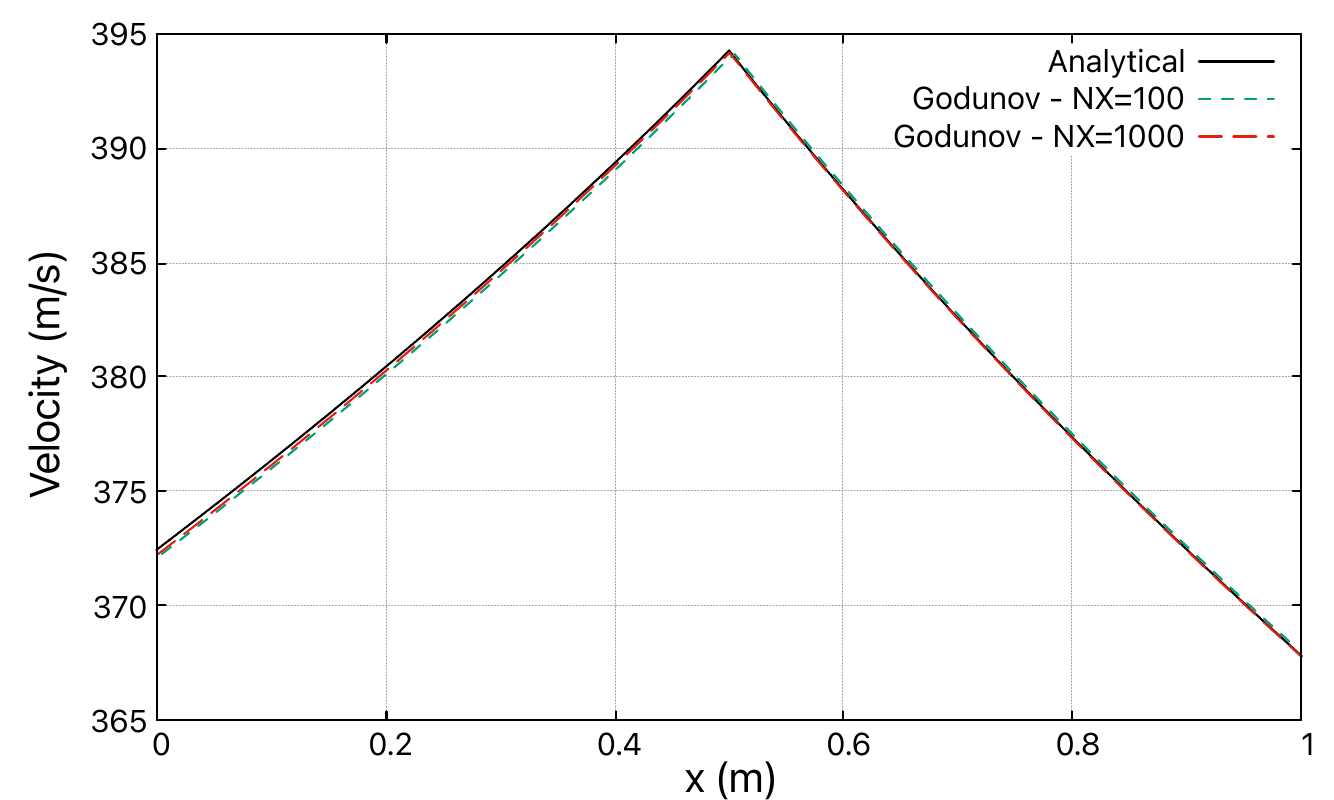}
        \label{fig:godu_varNX_sub_u_vapor}
    \end{subfigure}
    \caption{Analytical and numerical results comparison of a gas being heated and cooled down, for $\Pi > \Pi_1$, with mesh refinement.}
    \label{fig:godu_varNX_sub_vapor}
\end{figure*}

\begin{figure*}
\centering
    \begin{subfigure}{0.49\textwidth}
        \includegraphics[width=\textwidth]{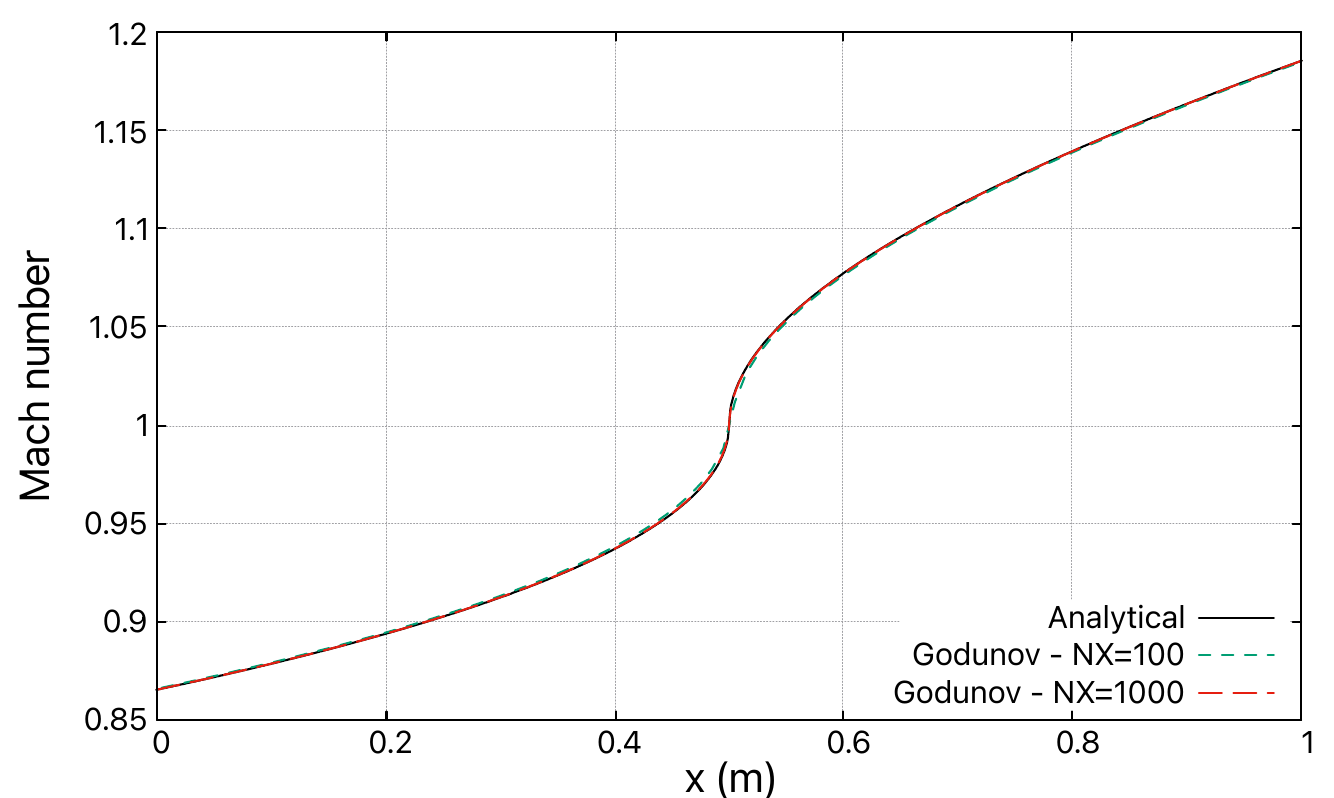}
        \label{fig:godu_varNX_sup_M_vapor}
    \end{subfigure}
    \begin{subfigure}{0.49\textwidth}
        \includegraphics[width=\textwidth]{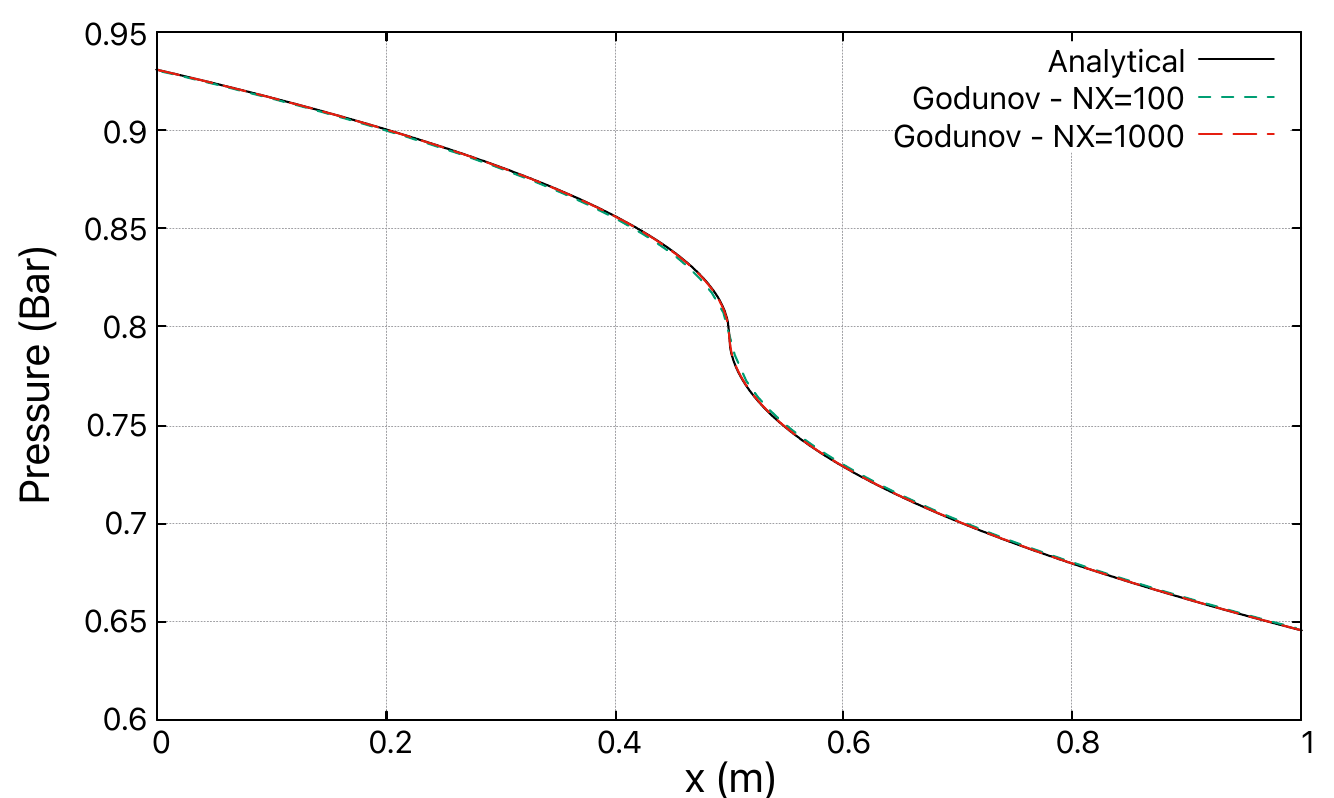}
        \label{fig:godu_varNX_sup_P_vapor}
    \end{subfigure}
    \begin{subfigure}{0.49\textwidth}
        \includegraphics[width=\textwidth]{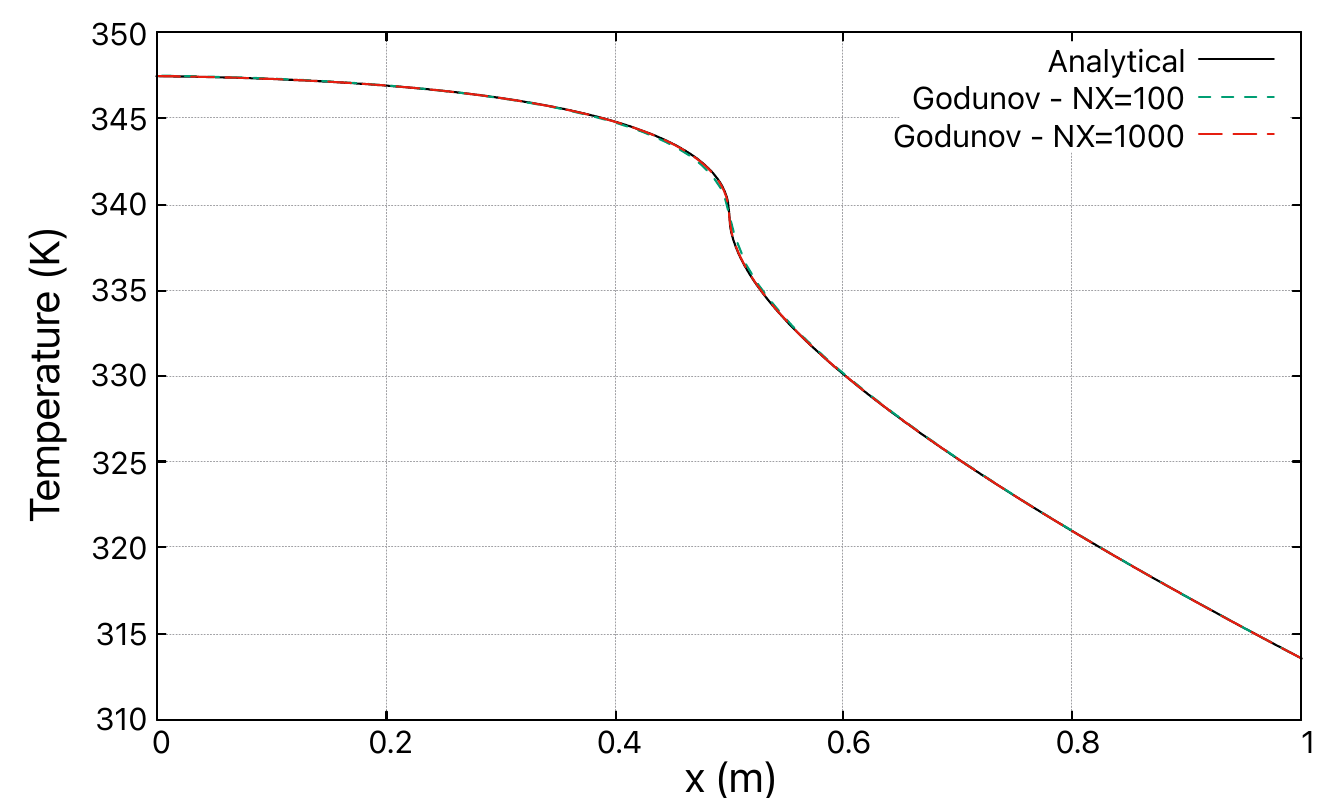}
        \label{fig:godu_varNX_sup_T_vapor}
    \end{subfigure}
    \begin{subfigure}{0.49\textwidth}
        \includegraphics[width=\textwidth]{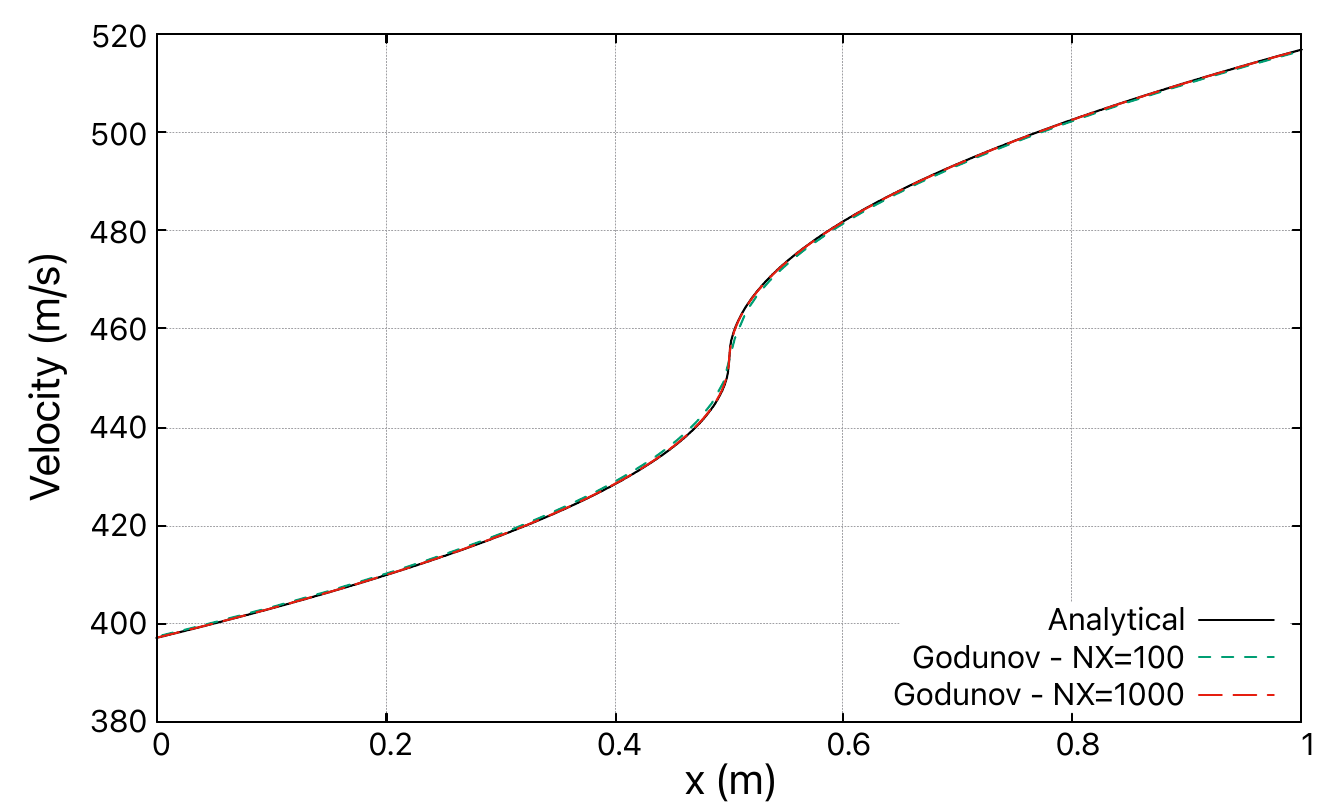}
        \label{fig:godu_varNX_sup_u_vapor}
    \end{subfigure}
    \caption{Analytical and numerical results comparison of a gas being heated and cooled down, for $\Pi < \Pi_1$, with mesh refinement.}
    \label{fig:godu_varNX_sup_vapor}
\end{figure*}

\begin{figure*}
\centering
    \begin{subfigure}{0.49\textwidth}
        \includegraphics[width=\textwidth]{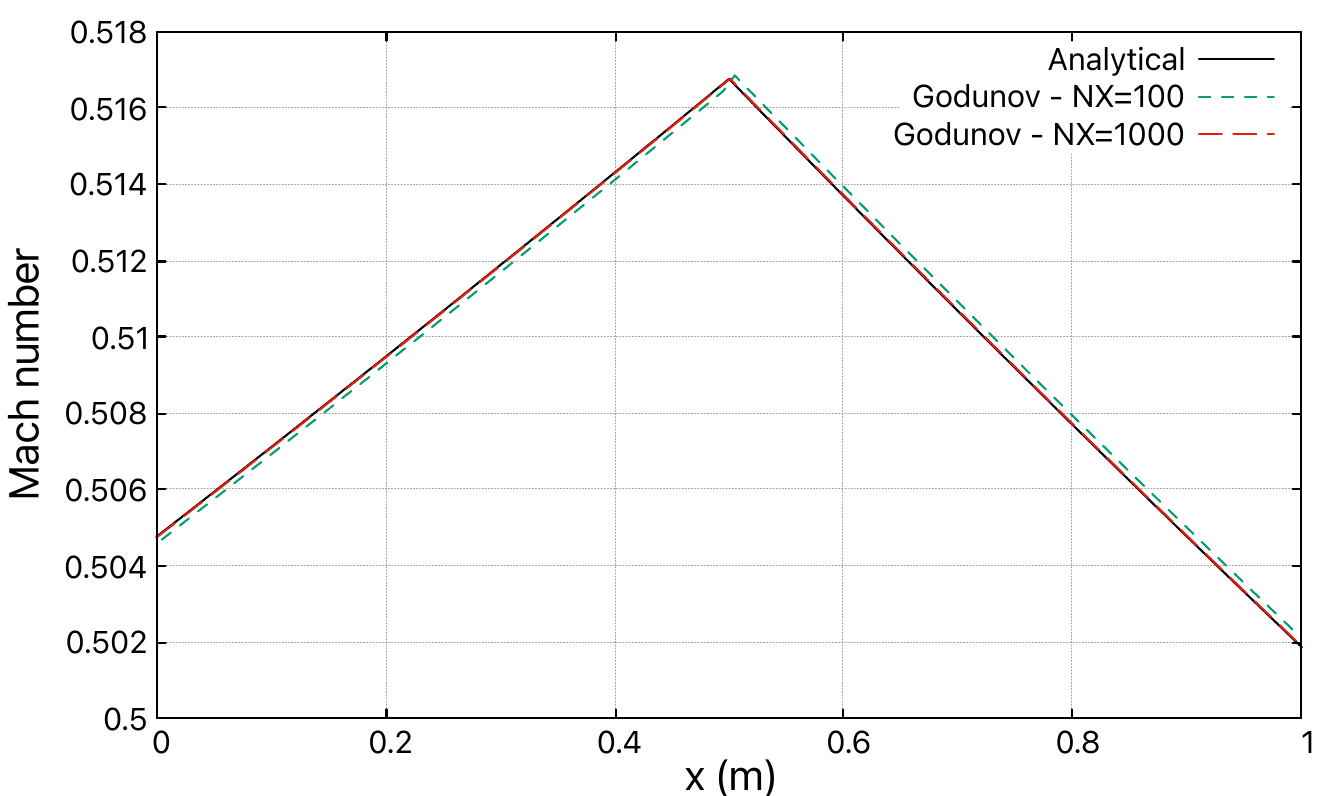}
        \label{fig:godu_varNX_sub_M_liquid}
    \end{subfigure}
    \begin{subfigure}{0.49\textwidth}
        \includegraphics[width=\textwidth]{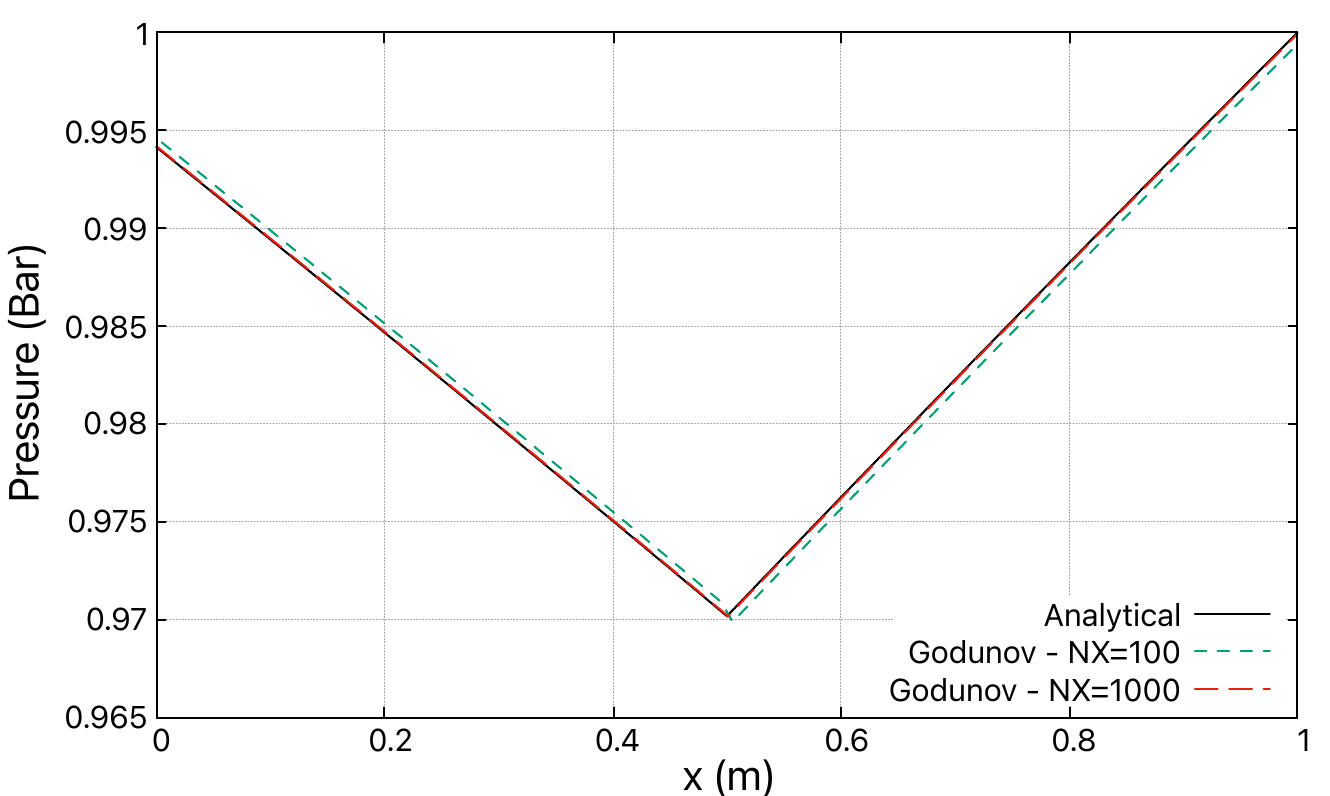}
        \label{fig:godu_varNX_sub_P_liquid}
    \end{subfigure}
    \begin{subfigure}{0.49\textwidth}
        \includegraphics[width=\textwidth]{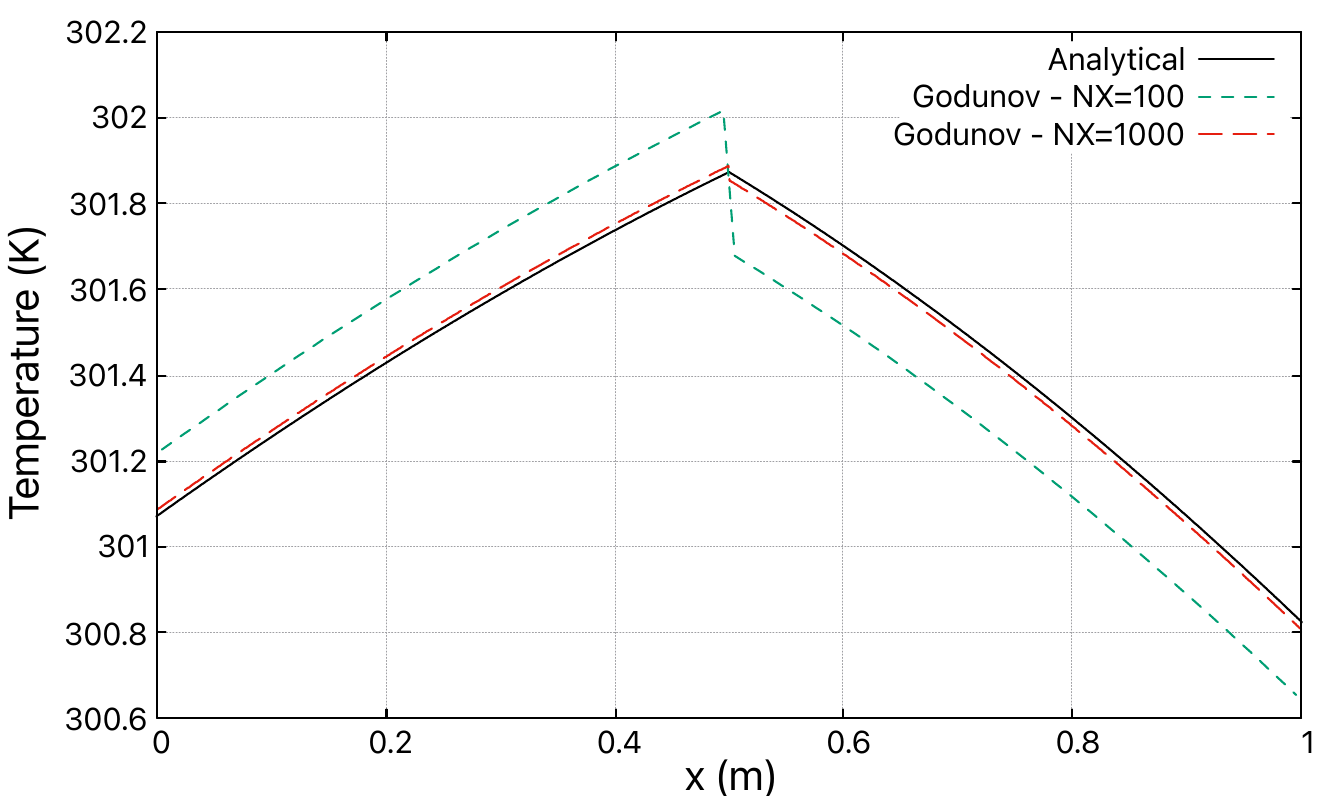}
        \label{fig:godu_varNX_sub_T_liquid}
    \end{subfigure}
    \begin{subfigure}{0.49\textwidth}
        \includegraphics[width=\textwidth]{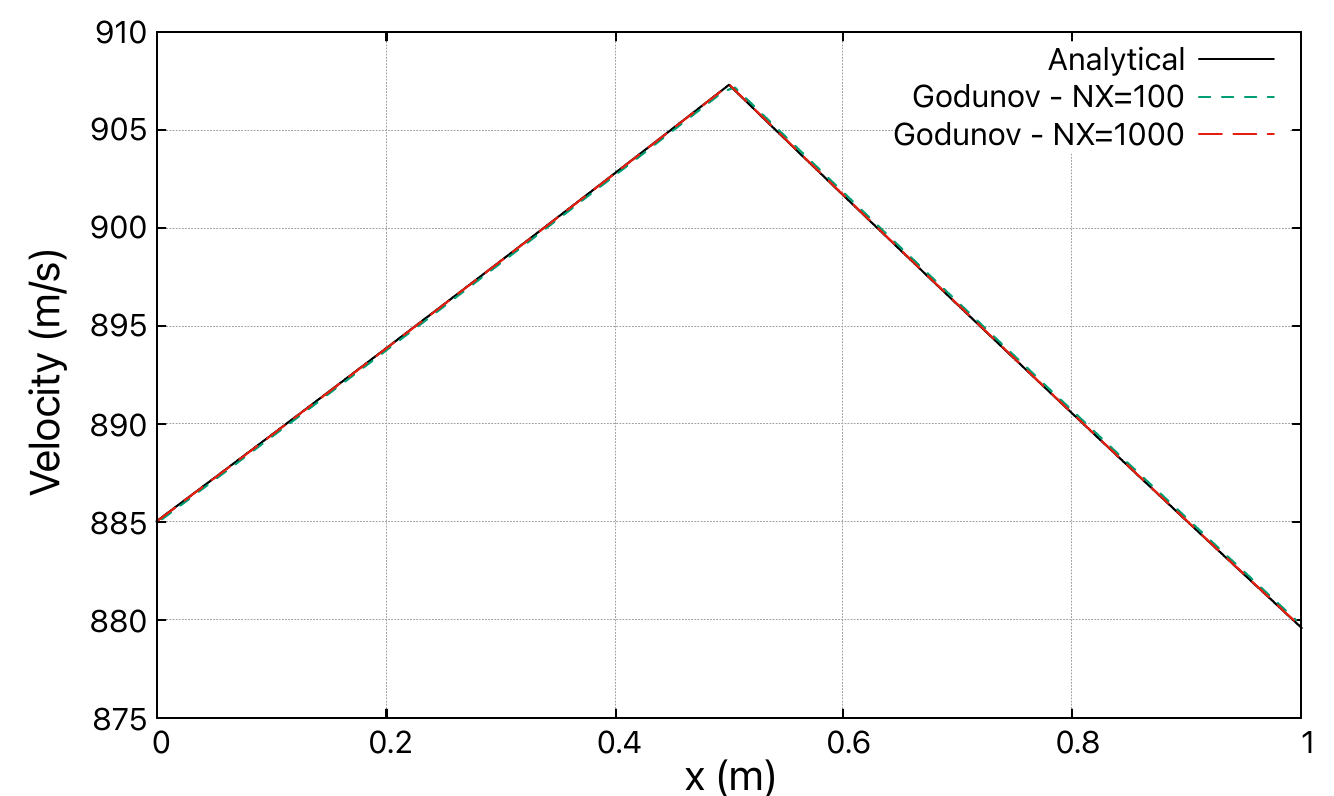}
        \label{fig:godu_varNX_sub_u_liquid}
    \end{subfigure}
    \caption{Analytical and numerical results comparison of a liquid being heated and cooled down, for $\Pi > \Pi_1$, with mesh refinement.}
    \label{fig:godu_varNX_sub_liquid}
\end{figure*}

\begin{figure*}
\centering
    \begin{subfigure}{0.49\textwidth}
        \includegraphics[width=\textwidth]{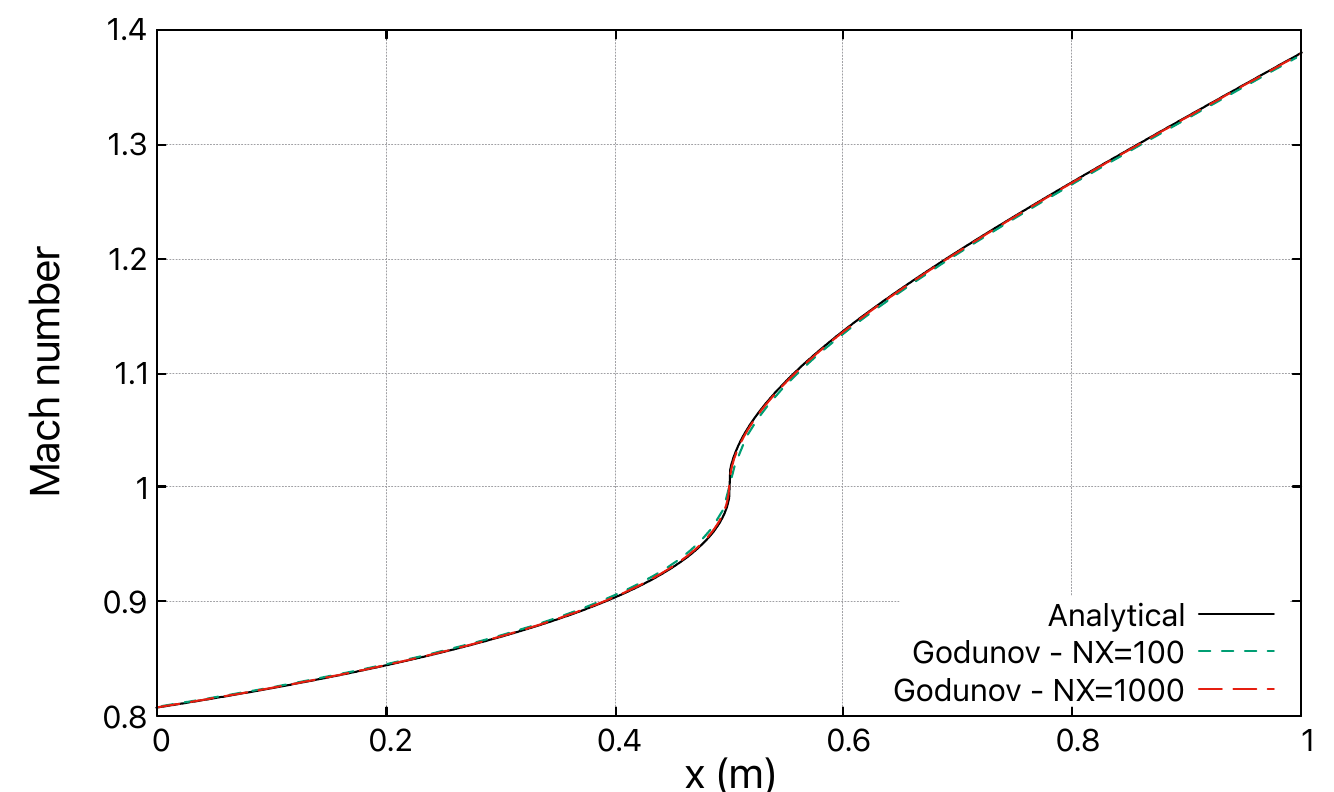}
        \label{fig:godu_varNX_sup_M_liquid}
    \end{subfigure}
    \begin{subfigure}{0.49\textwidth}
        \includegraphics[width=\textwidth]{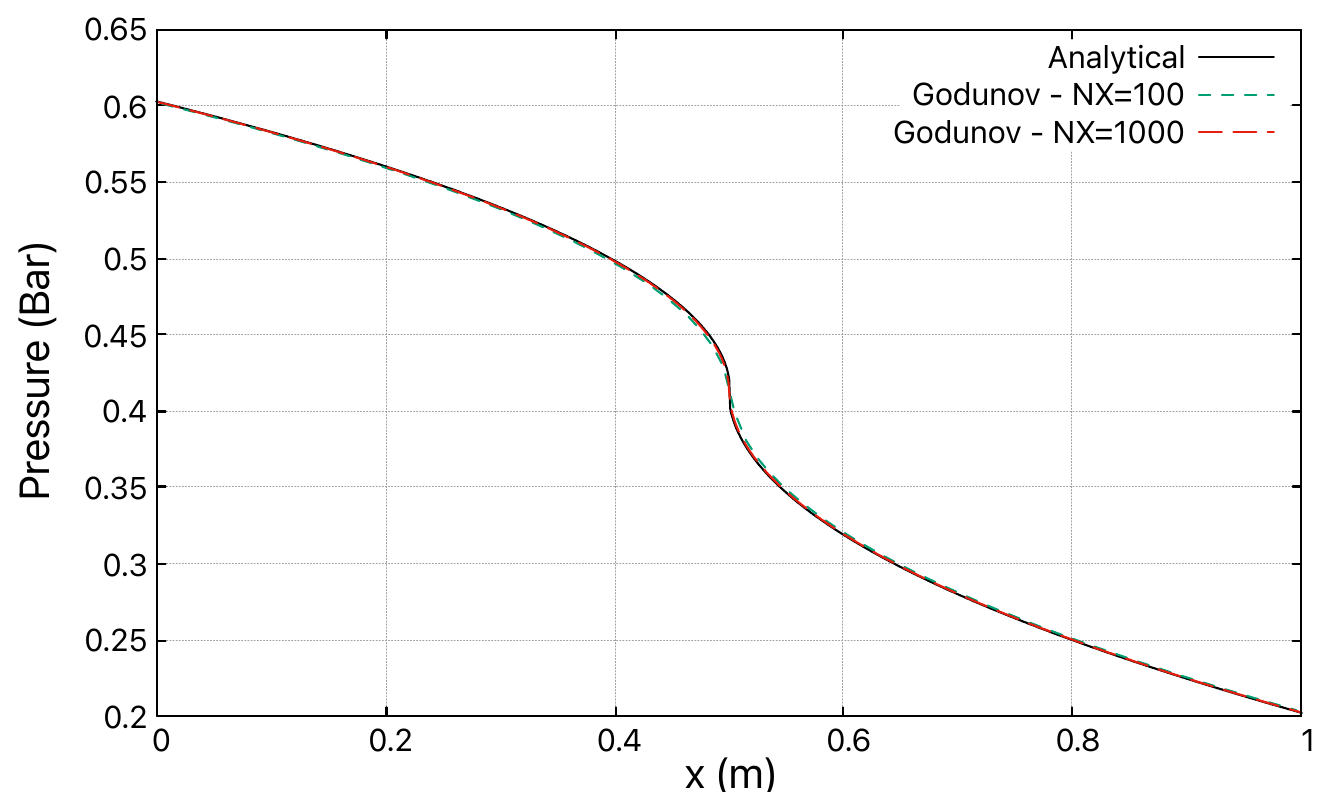}
        \label{fig:godu_varNX_sup_P_liquid}
    \end{subfigure}
    \begin{subfigure}{0.49\textwidth}
        \includegraphics[width=\textwidth]{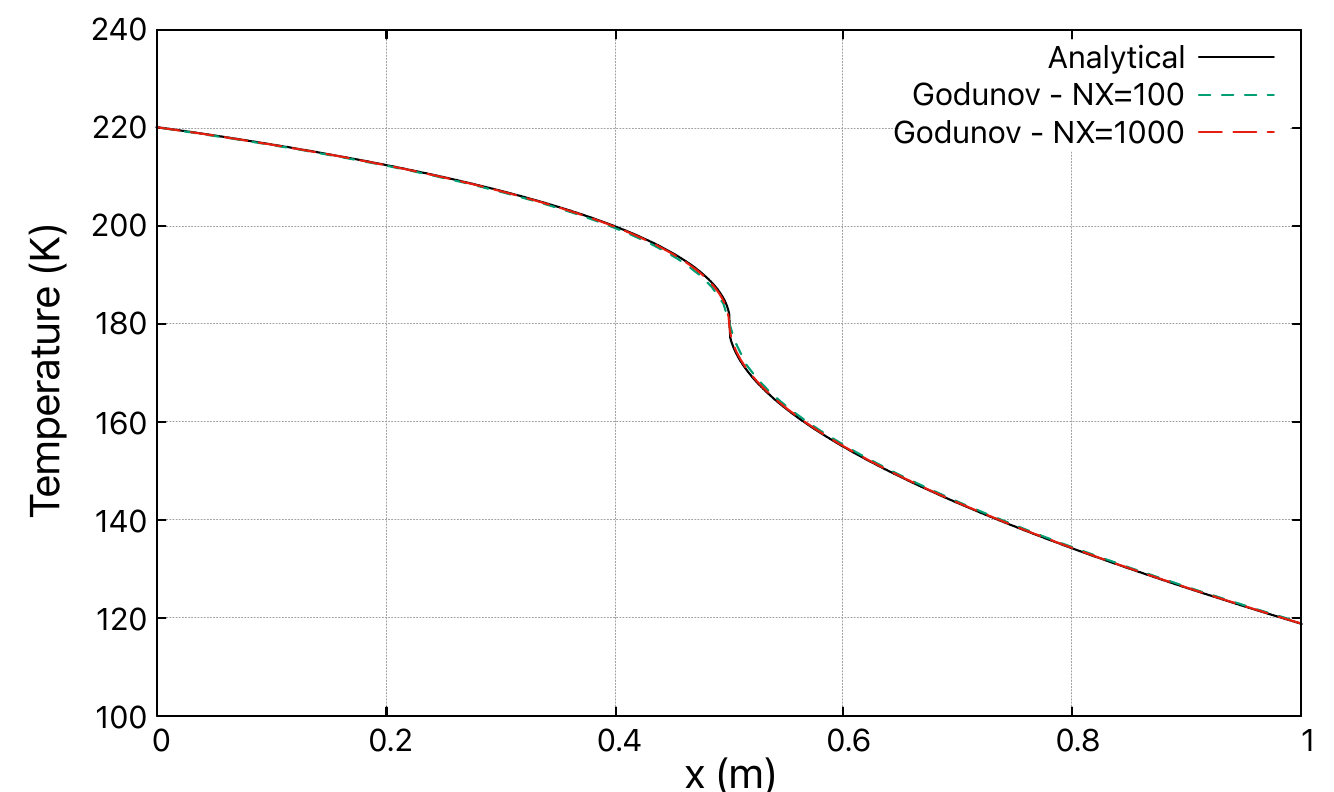}
        \label{fig:godu_varNX_sup_T_liquid}
    \end{subfigure}
    \begin{subfigure}{0.49\textwidth}
        \includegraphics[width=\textwidth]{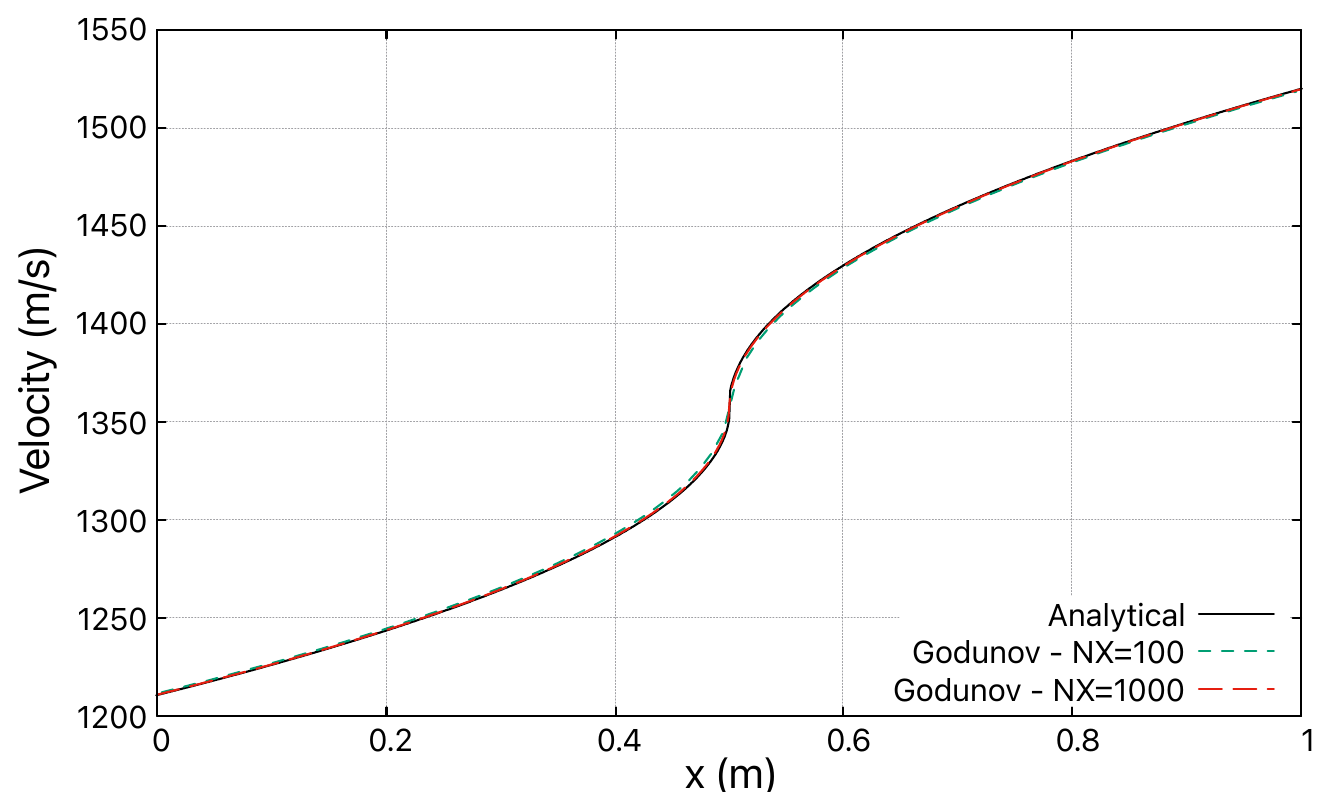}
        \label{fig:godu_varNX_sup_u_liquid}
    \end{subfigure}
    \caption{Analytical and numerical results comparison of a liquid being heated and cooled down, for $\Pi < \Pi_1$, with mesh refinement.}
    \label{fig:godu_varNX_sup_liquid}
\end{figure*}

For different levels of meshing (NX being the number of mesh cells), figure~\ref{fig:godu_varNX_sub_vapor} and figure~\ref{fig:godu_varNX_sub_liquid} compare the numerical code results with analytical fully subsonic results, respectively for the gas and the liquid. Figure~\ref{fig:godu_varNX_sup_vapor} and figure~\ref{fig:godu_varNX_sup_liquid} compare them with analytical supersonic outlet results, also respectively for the gas and the liquid.

We define the maximum relative error on variable $y$ as a function of meshing as follow: $\varepsilon_y = \text{max} \left( \frac{\vert y_{num} - y_{analytical} \vert }{ y_{analytical} } \right)$. Table~\ref{tab:numerical_error} displays the different maximum values of relative pressure error $\varepsilon_P$: corresponding to the previously mentioned figures, the finer the mesh, the closer the numerical solution gets to the analytical one. Moreover, when the solution is fully subsonic, the numerical solution is closer to the analytical one.

\begin{table}
    \begin{center}
    \def~{\hphantom{0}}
    \begin{tabular}{ c c c c c }
        \toprule
        {} &  \multicolumn{2}{ c }{Gas} & \multicolumn{2}{ c }{Liquid}\\
        \midrule
        $\varepsilon_P$   & $\Pi > \Pi_1$   & $\Pi < \Pi_1$    & $\Pi > \Pi_1$   & $\Pi < \Pi_1$\\
        
        NX = 100    &  0.106\% & 0.266\%   & 0.061\%  & 0.739\%\\
        NX = 1000   &  0.057\% & 0.085\%   & 0.006\%  & 0.222\%\\
        \bottomrule
    \end{tabular}
    \caption{Maximum values of relative pressure error depending on the level of meshing NX and the pressure ratio $\Pi$, for numerical simulations of a gas and a liquid.}
    \label{tab:numerical_error}
    \end{center}
\end{table}


    \section{Conclusion}
\label{sec:ccl}

Reference analytical or quasi-analytical solutions in heated and cooled rectilinear ducts have been addressed for compressible inviscid single-phase flows. They were developed by using the "Stiffened Gas" equation of state, allowing to describe gases or liquids.

Following the study done in a heated duct, a cooled subsection was added downstream to study its impact on the behavior of the flow. An analogy to nozzle behaviors was done through critical pressure ratios definition: the method to determine them was developed, depending on the value of the heat flux function applied to the duct and the tank conditions used to prescribe the inlet state.

The definition of two possible states was provided. When the pressure ratio between the outlet of the duct and the tank is greater than a specific critical pressure ratio, the mixture remains fully subsonic. When the pressure ratio becomes lower, the heated subsection reaches a sonic point at its end and the cooled subsection triggers a unique supersonic flow. The thermodynamic analysis showed that there cannot be a steady shock-wave in such flows.

The value of the critical pressure ratio demarcating the shift between the fully subsonic regime or the supersonic outlet regime was also studied: departing from the nozzle analogy where only a single ratio of areas is needed to determine the regime of the flow, both values of powers must be provided for heated and cooled ducts. Limit cases for this critical pressure ratios were developed, for cases without heat, extreme heating power or extreme cooling power.

Finally, results were provided, in good agreement with the analytical study. Varying the pressure ratio when it remains higher than the critical one leads to varying fully subsonic solutions with different mass flow rates. When it becomes lower, the cooled duct is fully supersonic and the solution is unique: the mass flow rate is fixed by the heated subsection and no more variation of solution happens. Locally, even with a more complex two-step function of power, varying the power applied results in predictable change of behavior described in \cite{shapiro1953dynamics}. 

The solutions stemming from this analytical study are expected to be used for validation of non-adiabatic compressible flows simulation tools. A short validation of a numerical code using the \cite{Godounov:1979ul} method was provided, yielding excellent results and very small relative errors.

To go further in this research work, one can expect that the numerical methods will be the favored method because of the complexity of the flow. Considering a non-adiabatic liquid flow, one can reasonably expect a mass transfer leading to a two-phase (liquid/vapor) flow. As we mentioned, some analytical solutions were also developed for two-phase flows in \cite{fathalli2018}: like the present paper, these studies provide help in analysing the flow as well allowing to check the quality of the codes by comparison with analytical solutions.

    \appendix

\appendixpage

\section{Solving $M_x$ from equation~(\ref{eqn:polyMach})}
\label{appA}

 $$
    f(M_x) = M_x^4 \left(\beta B - \gamma^2 A \frac{T_{0,x}}{T_{0,in}} \right) 
    + M_x^2 \left(B - 2\gamma A \frac{T_{0,x}}{T_{0,in}} \right)
    - A \frac{T_{0,x}}{T_{0,in}} = 0, 
    \begin{cases}
        \beta = \frac{\gamma-1}{2} \\
        A(P_{in}) = M_{in}^2 \left(1+\beta M_{in}^2 \right) \\
        B(P_{in}) = (1+\gamma M_{in}^2)^2
    \end{cases}
$$

Solving the previous equation yields 4 roots, 2 negatives which are not physical and 2 positives:
$$
    M_x(P_{in}) = \sqrt { \frac{ -b \pm \sqrt{\Delta} }{ 2a } },
    \begin{cases}
        a = \beta B - \gamma^2 A \frac{T_{0,x}}{T_{0,in}} \\
        b = B - 2\gamma A \frac{T_{0,x}}{T_{0,in}} \\
        c = -A \frac{T_{0,x}}{T_{0,in}} \\
        \Delta = b^2-4ac
    \end{cases}
$$

Thus we have the following positive Mach numbers:

$$
M_x^{sub}(P_{in}) =  \sqrt { \frac{ -b + \sqrt{\Delta} }{ 2a } } \leq 1
$$

$$
M_x^{sup}(P_{in}) =  \sqrt { \frac{ -b - \sqrt{\Delta} }{ 2a } } \geq 1 
$$

To be physical, there are two criteria to respect:

$$\Delta \geq 0$$
$$\frac{ -b \pm \sqrt{\Delta} }{ 2a } \geq 0$$


\section{Solving $Q_{cool}^{min}$ from equation~(\ref{eqn:Qcool_min})}
\label{appB}

The limit of our computation is $\Pi_1 = P_{out}/P_0 = 1$. Equation~(\ref{eqn:Pi}) is evaluated at the outlet for outlet pressure $P_{out}$ value:

$$
\Pi_1 = \frac{P_{out}}{P_0} = \frac{P_{in}^* + P_\infty}{P_0} \frac{1 + \gamma M_{in}^2}{1 + \gamma {M_{out}^{sub}}^2} - \frac{P_\infty}{P_0} = 1
$$

The isentropic relation~(\ref{eqn:stagnantpressure}) yields the following expression for inlet Mach number $M_{in}$:

$$
M_{in}^2 = \dfrac{2}{\gamma - 1} \left[ \left( \dfrac{P_0 + P_\infty}{P_{in}^* + P_\infty} \right) ^ {\frac{\gamma-1}{\gamma}}  - 1 \right]
$$

The lowest value of equation~(\ref{eqn:polyMach}) for a choked flow is evaluated at the outlet for $M_{out}^{sub}\left( \frac{T_{0,out}}{T_{0,in}} \right)$:

\begin{multline}
    {M_{out}^{sub}}^2 = \frac{ -b + \sqrt{\Delta} }{ 2a } = \frac{2\gamma \frac{T_{0,out}}{T_{0,in}}  - B + \sqrt{ \left( B - 2\gamma \frac{T_{0,out}}{T_{0,in}} \right)^2 + 4 A \frac{T_{0,out}}{T_{0,in}} \left(\beta B - \gamma^2 A \frac{T_{0,out}}{T_{0,in}} \right) }}{2 \left(\beta B - \gamma^2 A \frac{T_{0,out}}{T_{0,in}} \right)} , \\
    \begin{cases}
        \beta = \frac{\gamma-1}{2} \\
        A(P_{in}^*) = M_{in}^2 \left(1+\beta M_{in}^2 \right) \\
        B(P_{in}^*) = (1+\gamma M_{in}^2)^2
    \end{cases}\\
\end{multline}

The function $f \left( \frac{T_{0,out}}{T_{0,in}} \right) = \Pi_1 - 1 = 0$ is solved by a numerical method. Knowing the value of $\frac{T_{0,out}}{T_{0,in}}$, one can then evaluate $Q_{cool}^{min}$.
    
\printbibliography

\end{document}